\documentclass[prx,twocolumn,amsmath,amssymb,aps,longbibliography,superscriptaddress,citeautoscript,footinbib]{revtex4-2}

\usepackage{feynmp}
\usepackage{amsmath}
\usepackage{graphicx}
\usepackage{multirow}
\usepackage{latexsym}
\usepackage{textcomp}
\usepackage{verbatim}
\usepackage{color}
\usepackage{bm}
\usepackage{subfigure}

\usepackage{everysel}
\usepackage{keyval}
\usepackage{ragged2e}
\usepackage{dsfont}
\usepackage{amssymb}
\usepackage{enumitem}
\usepackage{pstricks}
\usepackage{mathtools}
\usepackage{kotex}

\usepackage{braket}
\usepackage{slashed} 
\usepackage{hyperref}        
\hypersetup{
     colorlinks=true,
     citecolor = red,    
     linkcolor=magenta,
     }

\usepackage{graphicx}
\usepackage{xcolor}
\usepackage{amsmath}
\usepackage{bbold}
\usepackage{amsfonts}
\usepackage{bbm}
\usepackage{comment}
\usepackage{orcidlink}
\usepackage[ruled,vlined]{algorithm2e}

\newcommand{\osum}{{%
    \setbox0\hbox{\circ}%
    \rlap{\hbox to \wd0{\hss\sum\hss}}\box0
}}

\newcommand{\mjh}[1]{\textcolor{teal}{#1}}

\begin{document}

\title{Frustrated superradiant phases in one- and two-dimensional lattices} 

\author{Jongjun M. Lee\,\orcidlink{0000-0002-9786-1901}}
\thanks{Contact author: jongjun@ualberta.ca}
\affiliation{Department of Physics and Quantum Horizons Alberta, University of Alberta, Edmonton, Alberta T6G 2E1, Canada}
\affiliation{Department of Physics, Pohang University of Science and Technology, Pohang 37673, Korea}

\author{Myung-Joong Hwang\,\orcidlink{0000-0002-0176-6740}}
\thanks{Contact author: myungjoong.hwang@duke.edu}
\affiliation{Division of Natural and Applied Sciences, Duke Kunshan University, Kunshan, Jiangsu 215300, China}

\begin{abstract}
Understanding how frustration and symmetry breaking shape collective behavior is a central problem in quantum many-body systems. In this work, we investigate this problem in large one- and two-dimensional arrays of coupled Dicke models on a periodic lattice, where strong light-matter coupling gives rise to a superradiant phase and competition between neighboring order parameters induces spontaneous translational symmetry breaking. Such Dicke lattice models constitute a fundamentally new class of quantum many-body systems, as they simultaneously realize the thermodynamic limit associated with the lattice size and an intrinsic thermodynamic limit arising from collective on-site interactions with quantum emitters. We show that frustration drives photonic density-wave ordering, and that the resulting broken periodicity can be predicted from the excitation spectrum of the symmetric phase, without requiring computationally prohibitive thermodynamic energy minimization. Furthermore, we demonstrate that an emergent Nambu-Goldstone mode arises near the critical point in a one-dimensional chain despite the presence of only discrete symmetry, and uncover the mechanism that enables this otherwise forbidden gapless excitation. We also find quasi-periodic ordering in the superradiant phase, reminiscent of quasicrystals, and demonstrate that synthetic magnetic flux provides a powerful knob to control the nature of translational symmetry breaking. Our results establish a new direction in quantum many-body physics where the coexistence of local and global thermodynamic limits gives rise to unconventional symmetry breaking and emergent collective behavior. 
\end{abstract}
 
\date{\today}
\maketitle

\section{Introduction}
Recent advances in quantum many-body physics have transformed strongly coupled light-matter systems into versatile platforms for investigating emergent collective phenomena and novel phases of matter~\cite{forndiaz2019ultrastrong,georgescu2014quantum}. Engineered quantum simulators, including trapped ions, ultracold atoms in optical cavities, and superconducting circuits, now enable unprecedented control over microscopic interactions, energy scales, and lattice geometries~\cite{blatt2012quantum,houck2012chip,ritsch2014cold,altman2021quantum}. Leveraging this control, experiments have realized paradigmatic models such as the Dicke and Rabi models, revealing superradiant phase transitions, dynamical critical behavior, and new forms of light-induced order~\cite{baumann2010dicke,klinder2015dynamical,hwang2015quantum,yoshihara2017superconducting,safavi2018verification,cai2021observation}. 

As research has progressed from single-mode cavities to extended arrays of coupled resonators, new questions have emerged regarding how spatial structure and competing ordering tendencies shape symmetry breaking and collective excitations in such systems~\cite{jin2013photon,zou2014implementation,hwang2016quantum}. In particular, geometric frustration, referring to situations in which local energy minimization is incompatible with global energy minimization~\cite{toulouse1987theory,moessner2006geometrical}, has recently attracted significant attention in various geometrically frustrated arrays of light-matter coupled systems, including Dicke~\cite{zhao2022frustrated,cheng2022chiral,zhao2023anomalous,zhang2024closed,luo2026quantum}, Rabi~\cite{zhang2021quantum,padilla2022understanding,li2023quantum,xu2024quantum,qin2024quantum}, Rabi models in ladder geometries~\cite{li2025meissner}, and Ising spins in ring geometries~\cite{Duan2023triangle}. Geometric frustration not only gives rise to rich quantum phases but also modifies quantum criticality, leading to anomalous scaling behavior. It also promotes competing orders and additional symmetry breaking, thereby enriching the collective many-body behavior of these systems.

However, this interplay of competing orders significantly increases the complexity of the problem, analogous to antiferromagnetic Ising spins on a triangular lattice~\cite{ramirez1994strongly}, where frustration leads to translational symmetry breaking~\cite{paddison2017continuous,bordelon2019field,bhattacharya2024evidence}. Most existing studies remain confined to a few sites or relatively small system sizes~\cite{zhao2022frustrated,cheng2022chiral,zhao2023anomalous,zhang2024closed,luo2026quantum,zhang2021quantum,padilla2022understanding,li2023quantum,xu2024quantum,qin2024quantum,jiang2025chirally,li2025meissner}, far from the thermodynamic limit of spatially extended systems. This limitation is especially pronounced in arrays of Dicke and Rabi models, where each site already possesses its own intrinsic thermodynamic limit~\cite{dicke1954coherence,hepp1973equilibrium,kirton2019introduction,ashhab2013superradiance,hwang2015quantum,hwang2016quantum}. In this case, reaching the thermodynamic limit of the array requires treating both the local thermodynamic limit at each site and the extensive limit across many sites~\cite{sachdev2011quantum} on equal footing. With these two coexisting thermodynamic limits, identifying the true ground state in general requires minimizing the energy over a landscape of competing orders whose complexity grows rapidly with lattice size, making the problem analytically and numerically challanging. As a result, the thermodynamic behavior of geometrically frustrated light-matter arrays remains largely unexplored, and addressing this challenge is the focus of the present work.

\begin{figure}[t!]
    \centering
    \includegraphics[width=\linewidth]{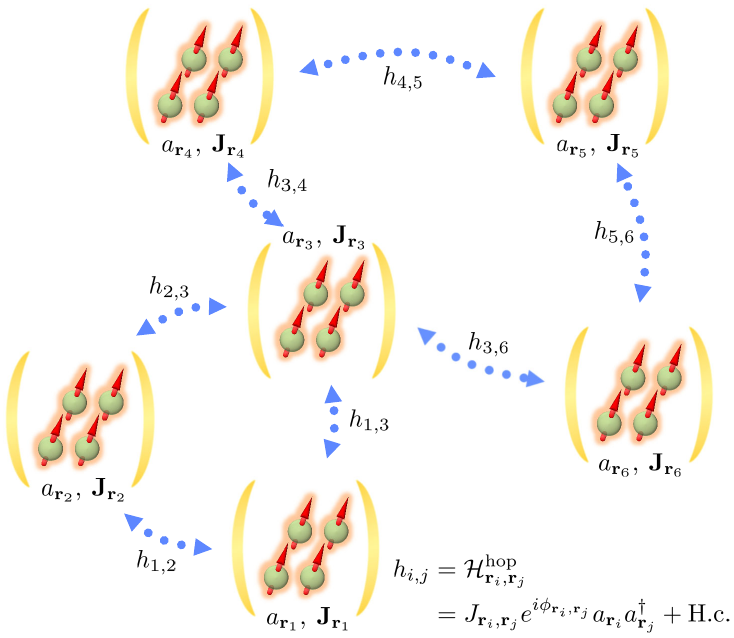}
    \caption{Schematic illustration of the Dicke lattice model. An example lattice that consists of six sites. Each node at $\mathbf{r}_{i}$ represents a Dicke model where a single cavity mode $a_{\mathbf{r}_{i}}$ is coupled to the collective spin $\mathbf{J}_{\mathbf{r}_{i}}$. The blue dotted arrow denotes the connectivity. $h_{i,j}$ denotes the hopping Hamiltonian $\mathcal{H}^{\rm hop}_{\mathbf{r}_{i},\mathbf{r}_{j}}$, which describes the photon hopping between site $\mathbf{r}_i$ and site $\mathbf{r}_j$ with the strength $J_{\mathbf{r}_{i},\mathbf{r}_{j}}$ and the phase $\phi_{\mathbf{r}_{i},\mathbf{r}_{j}}$.  }
    \label{fig1}
\end{figure}

Motivated by these challenges, we study superradiant phase transitions and geometric frustration in arrays of coupled Dicke models on spatially extensive periodic lattices, as schematically illustrated in Fig.~\ref{fig1}~\cite{dicke1954coherence,hepp1973equilibrium,kirton2019introduction}. In the superradiant phase, photon condensation accompanied by collective spin rotations leads to spontaneous breaking of translational symmetry~\cite{zhao2022frustrated,zhao2023anomalous}. We find that the excitation spectrum in the normal phase already encodes the reduced periodicity that emerges in the superradiant phase~\cite{luther1974single,hoesch2009giant}. Based on this insight, we develop a theoretical framework that enables us to treat translational symmetry breaking in large lattices. Within this framework, we investigate the ground-state configurations and excitation spectra in both the normal and superradiant phases.

The main results are as follows. We find that a Nambu-Goldstone mode emerges in the superradiant phase due to translational symmetry breaking in one-dimensional chains, but is present only in the vicinity of the critical point~\cite{nambu196dynamical,goldstone1962broken,watanabe2012unified}. This is striking because the underlying Hamiltonian possesses only discrete symmetry, which does not normally support a Nambu-Goldstone mode without any hidden continuous symmetry~\cite{Fan2014hidden,Leonard2017goldstone}. We uncover how the interplay between frustration and the two distinct thermodynamic limits allows the Landau theory to admit the Goldstone mode. Our results establish how the anomalous critical mode observed in small odd-site ring geometries~\cite{zhao2022frustrated,zhao2023anomalous,padilla2022understanding,li2023quantum,xu2024quantum} evolves in the thermodynamic limit, and show that even-odd parity continues to give rise to fundamentally distinct collective excitations in this limit. We further find that, on the triangular lattice, the superradiant phase exhibits (reduced) translational symmetry, in contrast to antiferromagnetic Ising spins on the same lattice~\cite{ramirez1994strongly}. We refer to this phase as a photonic density wave, by analogy with charge density wave order in condensed matter systems~\cite{gruner1988the,gruner1994the}. Finally, we show that the superradiant phase exhibits incommensurate order in ladder geometries, where the photon condensation generically develops a periodicity incommensurate with that of the underlying lattice~\cite{goldman1991quasicrystal,suck2013quasicrystals}. We demonstrate that this incommensurate periodicity can be tuned and restored to a commensurate one by introducing a magnetic flux. Although our presentation focuses on arrays of Dicke models, the framework equally applies to arrays of Rabi models~\cite{choi2020exotic,cai2021observation}.

This paper is structured as follows. In Sec.~\ref{Sec_pre}, we introduce the preliminaries, presenting the model in its general form and reviewing previous related studies. In Sec.~\ref{Sec_1D}, we present results and findings for the one-dimensional model, with a discussion on the emergent Nambu-Goldstone mode. In Sec.~\ref{Sec_2D}, we explore results for the two-dimensional model with the commensurate order on an infinite periodic lattice. In Sec.~\ref{Sec_Quasi}, we present findings for the quasi-one-dimensional model with both the incommensurate and the commrensure orders. In Sec.~\ref{Sec_Rabi}, we extend our results to arrays of Rabi models. In Sec.~\ref{Sec_Realization}, we discuss physical systems that can realize our model. Finally, we provide a summary in Sec.~\ref{Sec_Conclusion} and an outlook in Sec.~\ref{Sec_Outlook} for our findings.

\section{Preliminaries}\label{Sec_pre}
We consider an array of Dicke models with arbitrary connections, as shown in Fig.~\ref{fig1}. The total Hamiltonian is then comprised of both the local and hopping terms, expressed as follows.
\begin{equation}
\begin{aligned}
    H = \sum_{\bf r} \mathcal{H}^{(0)}_{\bf r} + \sum_{\mathbf{r},\mathbf{r}'} \mathcal{H}^{\rm hop}_{\mathbf{r},\mathbf{r}'},
\end{aligned}
\label{Eq_DA_Ham_0}
\end{equation}
where $\mathbf{r}$ indexes the position of each local site. The local term, $\mathcal{H}^{(0)}_{\mathbf{r}}$, describes a collection of two-level systems, such as spins or atoms, coupled to a single cavity mode, which reads
\begin{equation}
    \mathcal{H}^{(0)}_{\bf r} =  \omega_{\bf r} a^{\dagger}_{\mathbf{r}}a_{\mathbf{r}} + \Omega_{\bf r} J^{z}_{\mathbf{r}} 
+\frac{2\lambda_{\bf r}}{\sqrt{N_{a,\mathbf{r} }}}(a_{\mathbf{r}}+a^{\dagger}_{\mathbf{r}})J^{x}_{\mathbf{r}},
\label{Eq_H_loc_1}
\end{equation}
where $a_{\mathbf{r}}$ and $\mathbf{J}_{\mathbf{r}} = (J^{x}_{\mathbf{r}}, J^{y}_{\mathbf{r}}, J^{z}_{\mathbf{r}})$ represent the annihilation operator of the cavity photon and the collective total spin operator of size $N_{a}/2$ at $\mathbf{r}$, respectively. $\omega_{\mathbf{r}}$ and $\Omega_{\mathbf{r}}$ denote the frequencies of the cavity and the collective total spin in the $z$ direction, while $\lambda_{\mathbf{r}}$ represents the coupling strength between the cavity and the spin, and $N_{a,\mathbf{r}}$ indicates the number of microscopic spins with size $1/2$ at $\mathbf{r}$. $N_{a,\mathbf{r}}$ is given by a large number that governs the thermodynamic limit. The hopping part $\mathcal{H}^{\rm hop}_{\mathbf{r}, \mathbf{r}'}$ describes the hopping of the cavity photon between sites at $\mathbf{r}$ and $\mathbf{r}'$\mjh{,}
\begin{equation}
    \mathcal{H}^{\rm hop}_{\mathbf{r},\mathbf{r}'} = J_{\mathbf{r},\mathbf{r}'} e^{i\phi_{\mathbf{r},\mathbf{r}'}}a^{\dagger}_{\mathbf{r}'}a_{\mathbf{r}} +\text{H.c.},
\label{Eq_H_hop_1}
\end{equation}
where $\mathcal{H}^{\rm hop}_{\mathbf{r},\mathbf{r}'} = \mathcal{H}^{\rm hop}_{\mathbf{r}',\mathbf{r}}$, the effective magnetic flux is $\phi_{\mathbf{r},\mathbf{r}'} = -\phi_{\mathbf{r}',\mathbf{r}}$ and a real number, and the photon hopping energy $J_{\mathbf{r},\mathbf{r}'} = J_{\mathbf{r}',\mathbf{r}}$ is considered to be positive. The system exhibits a $\mathbb{Z}_2$ parity symmetry generated by
\begin{equation}
    \Pi = e^{i\pi \sum_{\bf r} \left(a^{\dagger}_{\mathbf{r}} a_{\mathbf{r}} + J^{z}_{\mathbf{r}} + N_{a,\mathbf{r}}/2 \right)},
\end{equation}
where the summation over $\mathbf{r}$ runs over all sites of the lattice. Under this symmetry, the operators transform as $a_{\mathbf{r}},\, J^{x,y}_{\mathbf{r}} \rightarrow -a_{\mathbf{r}},\, -J^{x,y}_{\mathbf{r}}$ for all $\mathbf{r}$, leaving the Hamiltonian invariant~\cite{zhao2022frustrated}. This can also be interpreted as a mirror reflection along the $z$ axis if the photon operator originates from a magnetic field of the cavity along the $x$ axis.

From now on, we adopt unit values for the distance between sites and for $\hbar$ to simplify the analysis. Our focus is to characterize ground state phases and the nature of quantum fluctuations in a closed system, neglecting dissipation due to environmental effects.
We assume a uniform Dicke lattice, where all on-site parameters are identical: $\omega_{\mathbf r}=\omega$, $\Omega_{\mathbf r}=\Omega$, $\lambda_{\mathbf r}=\lambda$, and $N_{a,\mathbf r}=N_{a}$ for all $\mathbf r$. The photon hopping is also taken to be uniform, $J_{\mathbf r,\mathbf r'} = J$ for all connected pairs.

The local part of the Hamiltonian in our model, Eq.~\eqref{Eq_H_loc_1}, corresponds to the standard Dicke model~\cite{dicke1954coherence,hepp1973equilibrium}. Although it lacks spatial degrees of freedom, a single cavity mode couples uniformly to $N_{a}$ two-level atoms and exhibits a superradiant phase transition in the thermodynamic limit $N_{a}\rightarrow\infty$~\cite{wang1973phase,kirton2019introduction}. In this limit, the mean-field theory becomes exact, since the cavity mode mediates an all-to-all interaction among the spins, and the thermodynamic limit is taken locally at each site. In the weak-coupling regime, the system remains in the normal phase, where the photon field is not condensed, and the collective spin aligns along $-\hat{z}$. As the light-matter coupling increases, the system undergoes a second-order phase transition at a critical point characterized by universal power-law scaling, signaling the onset of the superradiant phase~\cite{emary2003quantum,dimer2007proposed,bakemeier2012quantum}. This model thus provides a paradigmatic setting for quantum criticality in a zero-dimensional system with strong light-matter coupling.

Recently, considerable attention has been devoted to understanding how the superradiant phase transition is modified in spatially extended arrays, where a variety of novel phenomena have been identified, primarily in small or finite systems. For the well-known cavity QED models, several studies have investigated such effects in specific geometries:
\begin{enumerate}[label=(\arabic*)]
    \item{Two recent studies by one of the authors have examined closed Dicke arrays with a small odd number of identical sites ($N=3,5,7$) arranged in a ring geometry~\cite{zhao2022frustrated, zhao2023anomalous}, with tunable frustration controlled by the Peierls phase. They introduced the notion of frustrated superradiant phase transitions, whereby the geometric frustration of local superradiant order parameters leads to anomalous critical behaviors including the emergence of an additional critical mode with anomalous scaling, multicritical points, and asymmetric critical exponents. Further studies have explored variants of the Dicke trimer, including three-level atoms~\cite{cheng2022chiral}, unbalanced light-matter interactions~\cite{zhang2024closed}, and additional atom hopping interactions~\cite{luo2026quantum}, revealing rich critical phenomena due to frustrated superradiance.}

    \item{The role of the dissipation on the Dicke lattice model has also been investigated with a focus on the steady-state phase diagram. Ref.~\cite{zou2014implementation} has investigated a one-dimensional open Dicke array with negative hopping energy and the cavity damping. It demonstrates that a superradiant phase with spatially modulated order parameters, thus spontaneously breaking the translational symmetry, can be stabilized in the nonequilibrium steady state, with numerical solutions obtained for systems of up to around 100 sites. Ref.~\cite{xu2024phase,vivek2025self}, on the other hand, have investigated the two-site open Dicke models, known as the open Dicke dimer, demonstrating multistability across various phases. Moreover, recent studies on multiple spin ensembles in a single dissipative cavity has revealed nonreciprocal phase transitions~\cite{chiacchio2023nonreciprocal,lyu2025nonreciprocal}.}

    \item{Several recent studies have explored the Rabi lattice model in ring geometries, including triangular, square, and hexagonal configurations, both with and without an applied magnetic field~\cite{zhang2021quantum,padilla2022understanding,li2023quantum,xu2024quantum,qin2024quantum}. The ground-state phase diagram has been analyzed by mapping the system to spin models with exchange and Dzyaloshinskii-Moriya (DM) interactions~\cite{padilla2022understanding}, revealing that superradiant phases with chiral currents due to frustration can also arise in systems with an even number of sites~\cite{li2023quantum,qin2024quantum}. Moreover, the Rabi lattice model has been studied in a ladder geometry~\cite{li2025meissner}, where numerical results for systems of up to 16 sites demonstrate persistent chiral currents of cavity photons flowing along the edges, analogous to the Meissner effect in superconductors.}
    
    \item{A recent study has investigated a coupled spinning-top model, consisting of large Ising spins on a triangular geometry, where frustration is induced by antiferromagnetic spin-spin interactions~\cite{Duan2023triangle}. It reports a frustrated symmetry-broken phase accompanied by anomalous scaling behavior in the fluctuations of magnetic excitations.}

    \end{enumerate}
These studies highlight the rich physics of superradiant phase transitions on lattices, in both closed and open systems, which is only beginning to be revealed. However, they rely primarily on numerical optimization for relatively small system sizes, as accessing larger systems becomes prohibitively difficult due to the need to minimize a high-dimensional thermodynamic energy landscape.

\section{One-Dimensional Chain: emergence of Nambu-Goldstone mode}\label{Sec_1D} 
The one-dimensional Dicke chain with periodic boundary conditions has been studied as a minimal setting for exploring the superradiant phase transitions in lattice systems~\cite{zhao2022frustrated,cheng2022chiral,zhao2023anomalous,zhang2024closed,luo2026quantum,zhang2021quantum,padilla2022understanding,li2023quantum,xu2024quantum,qin2024quantum,jiang2025chirally,Duan2023triangle}. As the simplest periodic geometry, it already exhibits competition between local interactions and lattice structure, leading to novel phases due to geometrical frustration at the mean-field level~\cite{zhao2022frustrated,zhao2023anomalous}. 

In this section, we begin by briefly reviewing the model and its basic properties, and outlining the findings of previous studies along with the key outstanding theoretical challenges they pose. Numerical analyses based on small finite chains have revealed that the frustrated superradiant phases and the associated anomalous scaling exponents appear only in chains with an odd number of lattice sites, whereas even-site lattices exhibit a conventional superradiant phase. These findings suggest that the odd-site and even-site chain models belong to distinct universality classes and raise an important theoretical question about whether this even-odd parity dependence of the universality class persists in the thermodynamic limit of large lattices, where boundary effects are expected to be negligible. Addressing this question is the main aim of the present section, and we will show that the answer is affirmative, in striking contrast to the one-dimensional antiferromagnetic Ising model, where the even-odd parity does not affect the bulk properties in the thermodynamic limit.

\begin{figure}[t!]
    \centering
    \includegraphics[width=1.0\linewidth]{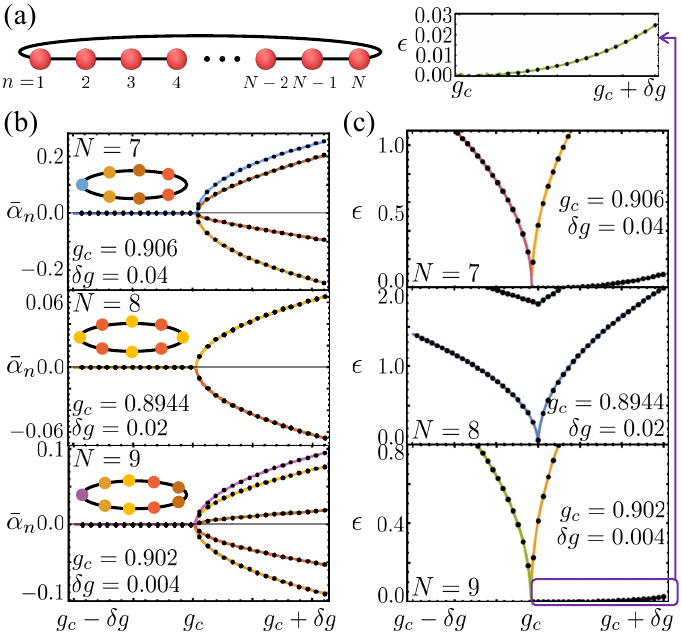}
    \caption{Schematic illustration and numerical results for the simple Dicke chain model. (a) A schematic of the model consisting of $N$ sites with the periodic boundary condition. Each red circle represents the Dicke model with a macroscopic number of spins and a single cavity mode. (b) The order parameter $\bar{\alpha}_n$ at site $n$ is plotted as a function of the effective coupling strength $g$, with $g_c$ denoting the critical effective coupling strength at which the superradiant phase transition occurs. The insets illustrate the spatial distribution in the superradiant phase in real space. (c) The excitation energies $\epsilon$ obtained by diagonalizing the quadratic Hamiltonian on a finite lattice in real space are shown as functions of the effective coupling strength $g$, and a low-lying mode of $N=9$ is highlighted in the purple box at a reduced energy scale. In (b) and (c), the black dots indicate numerical data, while the colored lines are guides to the eye. The parameters are $\omega=10$, $\Omega/\omega=1$ and $\bar{J}=0.1$ for all data.}
    \label{fig_1d_chain_data_1}
\end{figure}

Let us consider an array of identical Dicke models arranged along a one-dimensional chain of $N$ unit cells under periodic boundary conditions. Each unit cell contains a single Dicke model, and the unit cells are coupled through uniform nearest-neighbor hopping of the cavity photon, with a positive hopping energy $J_{\mathbf{r}, \mathbf{r}'} = J$. The magnetic flux is set to zero ($ \phi_{\mathbf{r}, \mathbf{r}'} = 0 $). We refer to this system as the simple Dicke chain model, which is schematically illustrated in Fig.~\ref{fig_1d_chain_data_1}(a). The Hamiltonian reads
\begin{equation}
\begin{aligned}
    H^{1d} =& \sum^{N}_{n=1} \Big[ \omega a^{\dagger}_{n}a_{n}+\Omega J^{z}_{n} + \frac{2\lambda}{\sqrt{N_{a}}} (a_{n}+a^{\dagger}_{n})J^{x}_{n}\\
    &+ J(a^{\dagger}_{n+1}a_{n}+a_{n+1}a^{\dagger}_{n} ) \Big],
\end{aligned}
\label{Eq_1d_Ham_1}
\end{equation}
where $n$ denotes the site index, and the remaining parameters are defined in the same way as in the general Hamiltonian from the previous section. The periodic boundary condition imposes the constraints $a_{n+N} = a_{n}$ and $J_{n+N}^{x,y,z} = J_{n}^{x,y,z}$. The Hamiltonian exhibits discrete translational symmetry, remaining invariant under the operation $n \rightarrow n+1$ for all indices, as well as $\mathbb{Z}_2$ parity symmetry.

The ground-state phase diagram of the Dicke chain can be obtained by using the mean-field theory. From the Hamiltonian in Eq.~(\ref{Eq_1d_Ham_1}), we write the renormalized mean-field energy with the renormalized parameters as follows.
\begin{equation}
\begin{aligned}
\bar{\mathcal{E}}^{1d} =& \sum^{N}_{n=1}\Big[ \bar{\alpha}^{2}_{n}+ \frac{1}{2}\cos\theta_{n} + g \bar{\alpha}_{n} \sin\theta_{n}\cos\phi_{n}\\
&+2\bar{J}\bar{\alpha}_{n}\bar{\alpha}_{n+1} \Big],
\end{aligned}
\label{Eq_1d_MF_1}
\end{equation}
where $\bar{\mathcal{E}}^{1d}=\langle H^{1d}\rangle /N_{a}\Omega$, $\bar{\alpha}_{n}=\sqrt{\omega/N_{a}\Omega }\langle a_{n}\rangle$ denotes the renormalized photon condensation as an order parameter, $g=2\lambda/\sqrt{\omega\Omega}$ denotes the effective coupling strength, $\bar{J}=J/\omega$, and $\theta_{n}$ and $\phi_{n}$ denote the spherical coordinates of the large spin at each position. 

In the weak coupling regime ($g\ll 1$), the large spin points along the $-z$ direction ($\mathbf{J}= -N_{a} \hat{z}/2$) with no photon condensation ($\alpha_{n}=0$) minimizes the mean-field energy. This corresponds to the normal phase, where the ground state preserves both the parity and discrete translational symmetries. However, in the strong coupling regime ($g\gg 1$), this trivial mean-field solution becomes unstable. The photon condenses ($\alpha_{n} \neq 0$), and the large spin rotates ($\langle \mathbf{J}_{n} \rangle \nparallel \hat{z}$), leading to the superradiant phase, where the parity is broken as in the single Dicke model. The critical coupling strength separating these regimes is given by $g_{c} = \sqrt{1+2\bar{J}\cos ( \frac{N-1}{N}\pi )}$ which approaches $\sqrt{1-2\bar{J}}$ in the limit $N\to \infty$~\cite{zhao2022frustrated}. The positive hopping interaction $J>0$, corresponding to the last term of the mean-field energy of Eq.~(\ref{Eq_1d_MF_1}), indicates that the order parameters prefer the alternating ordering of their signs, leading to the spontaneous breaking of the translational symmetry in the superradiant phase. 

To summarize the key features and consequences of the translational symmetry breaking in the Dicke chain model,  we present the numerical solutions for the ground-state order parameters for $N=7,8,$ and $9$ in Fig.~\ref{fig_1d_chain_data_1} (b). For even $N$, the order parameter configurations for $\{\alpha_n\}$ show the alternating signs with uniform magnitude, which is analogous to the Néel order in antiferromagnets and lowers the discrete translational symmetry by doubling the unit cell~\cite{kittel1996introduction}. For odd $N$, such Néel ordering is incompatible with the underlying lattice structure, resulting in a complete spontaneous breaking of translational symmetry due to geometric frustration. This is dubbed as a frustrated superradiant phase transition~\cite{zhao2022frustrated}. The different ways in which translational symmetry is broken in the order-parameter configurations for even and odd $N$, lead to correspondingly distinct quantum fluctuations around them~\cite{zhao2022frustrated,zhao2023anomalous,zhang2024closed,luo2026quantum}. These fluctuations can be analyzed by considering the quantum expansion around the mean-field solution using the Holstein-Primakoff transformation~\cite{holstein1940field}, which is described by a quadratic bosonic lattice Hamiltonian,
\begin{equation}
\begin{aligned}
H^{1d}_{\rm g} =& \sum^{N}_{n=1} \Big[ \omega a^{\dagger}_{n}a_{n}+\Omega \sqrt{1+4g^{2}\bar{\alpha}^{2}_{n}} b^{\dagger}_{n}b_{n} \\
&+ \lambda \frac{ \text{sgn}[\bar{\alpha}_{n}]}{\sqrt{1+4g^{2}\bar{\alpha}^{2}_{n}}} (a_{n}+a^{\dagger}_{n})(b_{n}+b^{\dagger}_{n})\\
&+ J(a^{\dagger}_{n+1}a_{n}+a_{n+1}a^{\dagger}_{n} ) \Big],
\end{aligned}
\label{Eq_1d_Ham_2}
\end{equation}
where $\bar{\alpha}_{n}$ is a solution that minimizes the mean-field energy [Eq.~(\ref{Eq_1d_MF_1})] and $b_{n}$ is the annihilation operator that describes the quantized fluctuation of the collective spin. As shown in Fig.~\ref{fig_1d_chain_data_1}(c), numerical diagonalization of the quadratic Hamiltonian on a finite lattice in real space gives the excitation energies, denoted by $\epsilon$. In the normal phase, where the Hamiltonian preserves translational symmetry, these eigenmodes can be associated with definite crystal momenta, although we do not explicitly label them by momentum in Fig.~\ref{fig_1d_chain_data_1}(c). The numerical results show that, for any $N$, there is a critical mode whose excitation energy follows a square-root behavior near the critical coupling strength, i.e., $ \epsilon \propto |g - g_c|^{1/2} $, a typical scaling for mean-field phase transitions. For odd numbers of sites, however, an additional critical mode emerges whose excitation energy deviates from the standard square-root behavior in the superradiant phase~\cite{zhao2022frustrated, zhao2023anomalous}, as shown in Fig.~\ref{fig_1d_chain_data_1}(c).
%[MJH:$\lambda_c$ expression is not given yet.] 
The anomalous critical properties, namely featuring two diverging critical time scales near the critical point, have been attributed to the geometric frustration of the order parameter, and has been found to be universal properties of the frustrated superradiant phases~\cite{zhao2022frustrated, zhao2023anomalous,fallas2022understanding,li2023quantum,Duan2023triangle,zhang2024closed,luo2026quantum}.

Numerical analysis for small system sizes up to $N=7$ suggested that the critical scaling of the additional frustrated mode, $\epsilon \propto |g - g_{c}|^{\gamma_{\rm F}}$, has the system-size dependent exponent $\gamma_{\rm F} = (N - 1)/2$~\cite{zhao2022frustrated}; see also Fig.~\ref{fig_1d_chain_data_1}(c). As we have already alluded to, this raises two important questions about the physics of the Dicke chain in the thermodynamic limit.  First, if $\gamma_{\rm F} = (N - 1)/2$ persists for large $N$, it implies the emergence of a massless excitation near the critical point $|g-g_c|\ll1$. Since the model contains no explicit continuous symmetry, this would appear incompatible with the well-known fact that massless (Goldstone) modes typically arise from spontaneous breaking of a continuous symmetry. Second, such massless excitation would survive in the thermodynamic limit only when $N$ is odd, and disappear when $N$ is even. Such a dramatic change of the bulk properties depending solely on the parity of $N$ suggests the presence of a delocalized excitation due to the geometric frustration. Insight into this scenario can be gained by comparing with the physics of 1D frustrated spin chains. In the 1D antiferromagnetic Ising spin chain, which has only a discrete $Z_{2}$ symmetry, the frustration induced by odd $N$ is localized at a single site, leading to identical bulk properties for any $N$. In contrast, in the 1D antiferromagnetic spin Heisenberg chain with continuous $U(1)$ symmetry, frustration for odd $N$ generates delocalized spinon excitations. While the Dicke chain possesses only a $Z_{2}$ symmetry, numerical evidence nevertheless suggests the possibility of a delocalized excitation analogous to those in the antiferromagnetic Heisenberg spin chain. 

Therefore, it is important to understand the nature of the order parameters in the large-$N$ limit and discover whether and how they may give rise to delocalized photon excitations induced by frustrated superradiance. The previous approaches based on numerical minimization of the mean-field energy and diagonalization of the quadratic bosonic modes are, however, not adequate to address this challenge. In the following, we therefore identify a theoretical framework that predicts the order parameters directly from the dispersion relation in the symmetric phase, and then use it to demonstrate the emergence of a massless excitation despite the underlying discrete $Z_{2}$ symmetry, uncovering the mechanism responsible for it.

\subsection{Critical modes and the translational symmetry breaking}
In the normal phase, the quadratic bosonic lattice Hamiltonian in Eq.~\eqref{Eq_1d_Ham_2} preserves the translational symmetry and therefore one can obtain a dispersion relation, namely, the dependence of the excitation energy on the crystal momentum $k$. Upon increasing the local qubit-oscillator interaction $\lambda$, the instability of the normal phase occurs whenever any one of the energies associated with a crystal momentum becomes zero. The normal mode corresponding to such a critical crystal momentum becomes macroscopically condensed. Therefore, while the crystal momentum becomes ill-defined in the superradiant phase in the presence of complete translational symmetry breaking as seen in Fig.~\ref{fig_1d_chain_data_1}(b), the critical crystal momentum would still determine the local order parameter distribution as it originates from the condensation of the corresponding normal mode in the vicinity of the critical point. This intuition is consistent with numerical observations showing that the order parameter follows a sinusoidal profile set by the critical crystal momentum~\cite{zou2014implementation}, and we formalize this connection and utilize it to predict the order parameters below.

\begin{figure*}[t!]
    \centering
    \includegraphics[width=0.89\linewidth]{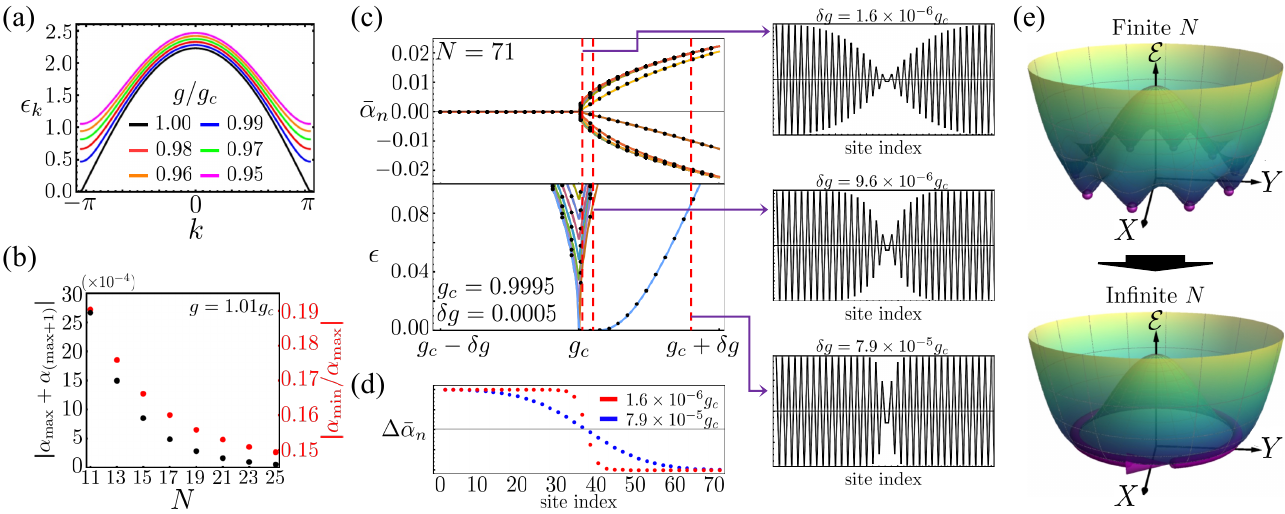}
    \caption{Numerical results for the simple Dicke chain model. (a) Excitation energy $\epsilon_{k}$ as a function of momentum $k$ for various coupling strengths in the normal phase, evaluated for a large system size. (b) Comparison of the order parameter across different system sizes. Black dots represent the amplitude difference between nearest neighbors, while red dots indicate the ratio between the maximum and minimum amplitudes of the order parameter, $\alpha_{\rm max}$ and $\alpha_{\rm min}$. For (a) and (b), the parameters are $\omega=10$, $\Omega/\omega=1$, and $\bar{J}=0.1$. (c) Order parameter and excitation energies $\epsilon$ as functions of the coupling strength for a chain of $71$ sites. The order parameter values are obtained by numerically minimizing the mean-field energy in Eq.~\eqref{Eq_1d_MF_1}, while the excitation energies are obtained from the real-space quadratic Hamiltonian. For selected coupling strengths marked by the dashed red lines, the real-space distribution of the order parameter is shown. Here, the black dots indicate numerical data, while the colored lines are guides to the eye. The parameters are $\omega=10$, $\Omega/\omega=1$, and $\bar{J}=0.0005$. (d) Renormalized Néel-vector amplitude $\Delta\bar{\alpha}_{n} = \bar{\alpha}_{2n} - \bar{\alpha}_{2n-1}$ plotted in real space over 71 sites. (e) Schematic illustration of the emergence of the Nambu-Goldstone mode in the simple Dicke chain model. For a finite chain, the mean-field energy $\mathcal{E}$ exhibits a discrete set of local minima (purple balls). In the infinite-chain limit, these minima merge into a continuous manifold, giving rise to a soft mode along the minimum-energy direction (purple arrow). Here, $X$ and $Y$ schematically denote two effective order-parameter components, corresponding to $\rho\cos\theta$ and $\rho\sin\theta$ in Eq.~\eqref{Eq_OP_2}, respectively. }
    \label{fig_1d_chain_data_2}
\end{figure*}

From  Eq.~\eqref{Eq_1d_Ham_2}, we derive the dispersion relation under periodic boundary conditions,
\begin{widetext}
\begin{equation}
\begin{aligned}
    \epsilon_{k}
    =\frac{1}{4}\sqrt{2\omega_{+}\omega_{-}+\xi^{2} -2\sqrt{ J^{2}\xi^{2} \cos^{2}{k}
    +2\xi^{2}(2\lambda^{2}-\omega_{+}\omega_{-})-4\omega_{+}(2\lambda^{2}-\omega\omega_{-})\xi+8\lambda^{2}\omega_{+}+\omega^{2}_{+}\omega^{2}_{-}
    }},
\end{aligned}
\label{Eq_disp_1}
\end{equation}
\end{widetext}
where $\omega_{\pm}=\Omega\pm\omega$, and $\xi=2\omega+J\cos{k}$. This dispersion relation is valid for both finite systems with discrete crystal momentum and the infinite system with continuous crystal momentum. Figure~\ref{fig_1d_chain_data_2}(a) shows the excitation energies for several coupling strengths $\lambda$, including the critical value. The results indicate that the modes with crystal momentum $\pm\pi$ become critical in the infinite-size limit. However, in the finite system, crystal momentum takes discrete values that depend on the system size. For systems with an even or odd number of sites $N$, the possible momentum values are given by,
\begin{equation}
k_m = \frac{2\pi m}{N},     
\end{equation}
where
\begin{equation}
m =
\begin{cases}
-\dfrac{N}{2}+1,\, \dots,\, \dfrac{N}{2}, & (N \ \text{even}), \\[6pt]
-\dfrac{N-1}{2},\, \dots,\, \dfrac{N-1}{2}, & (N \ \text{odd}).
\end{cases}
\end{equation}
For the case of an odd number of sites, the crystal momentum $k=\pm \pi$ is absent, with the nearest value being $k=\pm \frac{N-1}{N}\pi$. Following the dispersion relation in Eq.~(\ref{Eq_disp_1}), the momentum closest to $\pm \pi$ becomes the momentum of the critical mode, i.e.,
\begin{equation}
k_{c} =
\begin{cases}
\pi, & (N \ \text{even}), \\[6pt]
\pm \dfrac{N-1}{N}\pi, & (N \ \text{odd}).
\end{cases}
\label{Eq_discrete_k_1}
\end{equation}
Therefore, the critical mode $a_{k_c}$ that condenses in the superradiant phase, i.e., $\alpha_k = \langle a_k \rangle\neq0$ differs depending on whether the system has an even or odd number of sites.

Using the critical momentum, the real-space configuration of the order parameter near the critical point in the superradiant phase can be predicted. The local order parameter can be expressed as
\begin{equation}
    \alpha_{n} = \frac{1}{\sqrt{N}}\sum_{k} e^{ikn}\alpha_{k},
\end{equation}
where $1 \le n \le N$ is the site index, and $\alpha_k = \langle a_k \rangle = |\alpha_k| e^{i\theta_k}$, which follows from $a_k$ being the Fourier transform of $a_n$. For the case of even $N$, the term $\alpha_{\pi}$ becomes relevant in the summation for $\alpha_n$ due to its macroscopic condensation. Consequently, the order parameter in real space can be approximated as,
\begin{equation}
    \alpha_{n} \simeq \frac{1}{\sqrt{N}} |\alpha_{\pi}| \cos(\pi n) \propto (-1)^{n},
\label{Eq_alpha_even_1}
\end{equation}
where $\theta_{\pi} = 0$ is set due to the gauge choice. This result indeed confirms the halving of the reduced translational symmetry or the Néel order in the superradiant phase as shown in Fig.~\ref{fig_1d_chain_data_1}(b) and agrees with numerical results for any large N (not shown).

In contrast, for odd $N$, there are two degenerate critical Fourier modes at $k_c = \pm \frac{N-1}{N}\pi = \pm\left(\pi - \frac{\pi}{N}\right)$, which lie slightly away from the Brillouin-zone edge. Both critical components contribute to the formation of the real-space order parameter. Retaining only these two critical contributions in the summation for $\alpha_n$ and using the reality condition $\alpha_{-k}=\alpha_k^{*}$, the real-space order parameter takes the form
\begin{equation}
\begin{aligned}
\alpha_n 
&\simeq \frac{2}{\sqrt{N}} |\alpha_{k_c}|
\cos\!\left(\frac{N-1}{N}\pi n\right) \\
&= \frac{2}{\sqrt{N}} |\alpha_{k_c}|\, (-1)^n
\cos\!\left(\frac{\pi}{N} n\right).
\end{aligned}
\label{Eq_alpha_odd_1}
\end{equation}
This expression reveals a staggered order parameter consisting of a short-wavelength staggered component with wavevector $\pi$, multiplied by a long-wavelength modulation of wavelength $2N$. The latter originates from the fact that the critical mode lies slightly away from the Brillouin-zone boundary.

As shown in the top-right panel of Fig.~\ref{fig_1d_chain_data_2}(c) with $N=71$, this analytical form given in Eq.~\eqref{Eq_alpha_odd_1} agrees well with the numerical data of the local order parameter obtained by numerically minimizing the mean-field energy in Eq.~\eqref{Eq_1d_MF_1} in the vicinity of the critical point, thereby clarifying the microscopic origin of the complete breaking of translational symmetry for odd $N$. Eqs.~\eqref{Eq_alpha_even_1} and \eqref{Eq_alpha_odd_1} highlight that changing the system size by a single site globally alters the order parameters. For even $N$, the allowed momenta include the zone-boundary mode at $k=\pi$, whereas for odd $N$ this mode is absent and the instability instead occurs at momenta shifted from the zone boundary by an amount of order $1/N$. This seemingly minor shift in the critical momentum has a profound consequence in real space: it changes the order parameter from a purely staggered configuration into one that is modulated by a long-wavelength envelope extending over the entire system. Consequently, the distinction between even and odd $N$ is not a local effect but a global one that affects the bulk property.

We note that the above method of estimating the order-parameter distribution based on the critical momentum can substantially reduce the computational cost of solving the mean-field solutions for any large lattices with translational symmetry, compared with the direct minimization of the mean-field energy. The algorithm for obtaining the mean-field solution is presented in Appendix~\ref{Appx_Algorithm}, and the application of this method to the Dicke ladder model is discussed in detail in Sec.~\ref{Sec_Quasi}.

\subsection{Nambu-Goldstone mode and domain wall}
Having established that the superradiant order parameter for odd $N$ takes the form of a staggered configuration modulated by a long-wavelength envelope, we now examine the implications of this structure for the low-energy excitation spectrum. As shown in previous sections, the odd-$N$ Dicke chain hosts an additional soft mode, as one of the two critical modes, whose critical exponent increases proportionally to $N$. These two closely linked properties are clearly visible in Fig.~\ref{fig_1d_chain_data_2}(c), which presents the spatial profile of the order parameter at representative coupling strengths and the excitation spectrum for a chain of $N=71$, significantly larger than system sizes explored in previous studies. As $N$ increases, the energy of the anomalous mode is progressively suppressed and approaches zero in the vicinity of the critical point. Near criticality, where this mode remains soft, the numerically obtained order parameter is accurately captured by the analytical expression in Eq.~(\ref{Eq_alpha_odd_1}). Away from the critical point, once the order parameter deviates from this form by gradually shrinking the range of sites where the amplitude modulation of the staggered configuration occurs, the anomalous mode acquires a finite mass.

We attribute the origin of this anomalous soft mode and understand its fate in the thermodynamic limit from the analytical structure of the order parameter, Eq.~(\ref{Eq_alpha_odd_1}). Note that a translation by two lattice sites ($\mathcal{T}_2 $) transforms the order parameter as
\begin{equation}
\mathcal{T}_2 \alpha_n \equiv \alpha_{n+2}
= \alpha_n\left(\theta + \delta\theta\right),
\: \delta\theta = \frac{2\pi}{N},
\label{Eq_transl_op_1}
\end{equation}
As the system size increases, $\delta\theta \to 0$, implying that the discrete translation operation continuously shifts the phase of the long-wavelength envelope. In the large-$N$ limit, therefore, all $N$ symmetry-broken order-parameter configurations become continuously connected through a continuous phase shift, and the minima of the mean-field energy form an effectively continuous manifold reminiscent of a Mexican-hat potential. See Fig.~\ref{fig_1d_chain_data_2}(e). This emergent continuous degeneracy gives rise to a collective excitation whose energy vanishes asymptotically with increasing $N$. We therefore identify this mode as an emergent Nambu-Goldstone mode associated with the spontaneous breaking of translational symmetry in the odd-$N$ superradiant phase~\cite{nambu196dynamical,goldstone1962broken,watanabe2012unified}.

The nodes in Fig.~\ref{fig_1d_chain_data_2}(c) are the locations where the Néel ordering pattern of alternating signs changes orientation, from a plus-minus arrangement on one side to a minus-plus arrangement on the other. In this sense, the node acts as the center of the domain wall. Near the critical point, the domain-wall size is extensive, scaling as $O(N)$; in this regime, the associated Nambu-Goldstone mode implies that the energy cost to shift the domain wall by a single site is negligible. Far from the critical point, however, the order parameter no longer follows Eq.~(\ref{Eq_alpha_odd_1}), and the domain wall becomes sharply localized with a width of order $O(1)$, independent of $N$. Consequently, the order parameter becomes increasingly flat away from the domain wall as the system size grows, and although the amplitude at the wall increases compared to the near-critical regime, it nevertheless decreases toward zero with increasing $N$; see Figs.~\ref{fig_1d_chain_data_2}(b) and (c). In this non-critical regime, the order-parameter configurations are no longer smoothly connected by translation operations, so shifting the domain wall requires a finite energy, giving the anomalous mode a finite excitation gap. The domain-wall structure and its width are more clearly visible in the Néel vector, defined as the difference between neighboring order-parameter values, as shown in Fig.~\ref{fig_1d_chain_data_2}(d).

We now formalize the emergence of a Mexican-hat potential from a Hamiltonian only with discrete symmetry within a Ginzburg-Landau framework, establishing it as a general mechanism through which gapless low-lying excitations arise in the absence of a continuous symmetry. Near the critical point, the mean-field energy in Eq.~\eqref{Eq_1d_MF_1} can be expanded as
\begin{equation}
\bar{\mathcal{E}} \simeq \sum^{N}_{n=1}\sum^{\infty}_{j=0}c_{2j}\bar{\alpha}^{2j}_{n} + 2\bar{J}\sum^{N}_{n=1}\bar{\alpha}_{n}\bar{\alpha}_{n+1},
\end{equation}
where $N$ is an odd number and $c_{0}=-1/2$, $c_{2}=1-g^{2}$, $c_{4}=g^{4}$, and so on. The order parameter is approximately given by
\begin{equation}
\bar{\alpha}_{n} \simeq \rho \cos (k_{c}n+\theta)=\cos(k_{c}n)X-\sin(k_{c}n)Y,
\label{Eq_OP_2}
\end{equation}
where $\rho$ is the amplitude, $\theta$ is the phase, $X=\rho\cos\theta$, and $Y=\rho\sin\theta$. In this representation, the amplitude and phase can be viewed as polar coordinates in an effective two-dimensional order-parameter space. The corresponding Cartesian components, denoted schematically by $X$ and $Y$ in Fig.~\ref{fig_1d_chain_data_2}(e), parametrize the low-energy manifold of symmetry-broken configurations. Substituting this ansatz into the mean-field energy, the amplitude is determined by the quadratic and quartic terms, while the phase dependence first appears at order $2N$. With the value of critical momenta $k_{c}=\pi (N-1)/N$ in Eq.~\eqref{Eq_discrete_k_1}, the energy can thus be written as
\begin{equation}
\bar{\mathcal{E}} \simeq \bar{c}_{2}\rho^{2}+ \bar{c}_{4}\rho^{4}+  \cdots+ \bar{c}_{2N}\rho^{2N}\cos(2N\theta)+\cdots,
\label{Eq_Expand_MF_1}
\end{equation}
where $\bar{c}_{j}$ depends on $N$. We emphasize that all contributions below order $2N$ are independent of $\theta$. Minimizing the energy with respect to $\rho$ yields the stable amplitude $\rho_{0} \propto \sqrt{g-g_{c}}$~\cite{zhao2022frustrated}. The phase is determined from $\partial_{\theta}\bar{\mathcal{E}}=0$ and $\partial^{2}_{\theta}\bar{\mathcal{E}}>0$, giving $\theta_{0} = m\pi/N$ where $m$ is an integer and $\bar{c}_{2N}<0$. Expanding the energy around the mean-field solution, one obtains
\begin{equation}
\bar{\mathcal{E}} = 4\bar{c}_{4}\rho^{2}_{0} (\delta \rho)^{2} -2 \bar{c}_{2N}N^{2}\rho^{2N-2}_{0} (\rho_{0}\delta \theta)^{2} + \cdots.
\end{equation}
Here, the phase fluctuation carries an additional factor of $\rho_{0}$ due to the metric of the angular coordinate. Consequently, the amplitude mode exhibits the conventional scaling $\omega_{\rho} \propto \sqrt{g-g_{c}}$, while the phase mode shows an anomalous scaling $\omega_{\perp} \propto (g-g_c)^{(N-1)/2}$ leading to the low-lying excitation. This anomalous behavior originates from the translational symmetry, which forbids any phase-dependent contribution below order $2N$ and thereby strongly suppresses the stiffness of the angular mode. As noted above, this analysis is valid only near the critical point.

The expansion of the mean-field energy in Eq.~\eqref{Eq_Expand_MF_1} provides another way to understand the emergent continuous degeneracy discussed above. Near the critical point in the superradiant phase, the low-order terms, such as $\rho^{2}$ and $\rho^{4}$, dominate the energy and are independent of the phase $\theta$. The first $\theta$-dependent contribution appears only at order $\rho^{2N}$, which is strongly suppressed near criticality for large $N$. Therefore, the mean-field energy becomes approximately invariant under a continuous shift of $\theta$, giving rise to an emergent $U(1)$-like structure along the angular direction [Fig.~\ref{fig_1d_chain_data_2}(e)]. As the system moves away from the critical point, higher-order $\theta$-dependent terms become increasingly important, lifting this approximate degeneracy and giving a finite gap to the phase mode. This picture is consistent with the translation argument in Eq.~\eqref{Eq_transl_op_1}.

Our analysis captures the evolution of the anomalous soft mode observed in small chains to the emergent Nambu-Goldstone mode in the large-size limit. We refer to this as an emergent Nambu-Goldstone mode because it appears only (1) near the critical point and (2) in sufficiently large systems. The emergent Nambu-Goldstone mode in the Dicke chain is analogous to acoustic phonons in periodic crystals and to phasons in charge-density-wave phases, both of which are gapless collective excitations arising from the spontaneous breaking of translational symmetry in condensed-matter systems~\cite{cowley1976acoustic,gruner1988the}. The phason is particularly relevant to our case, as it corresponds to the spontaneous breaking of discrete lattice translational symmetry. It is expected to remain massless as a Nambu-Goldstone mode; however, recent studies have shown that long-range Coulomb interactions can render the phason massive by lifting its gaplessness~\cite{anderson1963plasmons,virosztek1993collective,kim2023observation}. In contrast, our emergent Nambu-Goldstone mode is gapless only near the critical point and becomes massive away from the critical point. Here, the gap is lifted by moving away from criticality itself, rather than by the presence of long-range interactions.

Finally, we emphasize that the evolution of the Nambu-Goldstone mode with system size has not been explored previously, as spontaneous symmetry breaking is usually discussed only after taking the thermodynamic limit in which the lattice size itself diverges. In contrast, the Dicke lattice model admits an intrinsically defined thermodynamic limit at the level of each cavity, associated with the large collective spin (or large atom number) limit. This allows spontaneous symmetry breaking to occur already for a finite lattice. Increasing the number of lattice sites $N$ therefore does not merely introduce finite-size effects but instead controls an independent limit that governs the emergence of spatially collective excitations. As a result, the Nambu-Goldstone mode identified here emerges in a qualitatively distinct manner, becoming asymptotically massless as $N$ increases, with its origin rooted in the frustration of the order parameter.

\subsection{Bulk properties away from criticality}
The preceding analysis demonstrates that the presence of a Nambu-Goldstone mode for odd $N$, absent for even $N$, arises from the extensive nature of the domain wall. Away from the critical point, however, the size of the domain wall becomes $O(1)$ and such a local defect is not expected to affect the bulk properties in the large-$N
$ limit. In fact, the order parameter configuration away from the domain wall becomes nearly Néel ordered even for odd-$N$. Consequently, far from both the critical point and the domain wall, the system can be effectively described by a theory with a commensurate order, in which the periodicity of the order parameter is reduced by half and remains commensurate with the underlying lattice. This analysis is analogous to magnetic systems, where domain walls appear as localized defects embedded in an otherwise periodic order, and spin excitations are typically examined against such commensurate backgrounds~\cite{blundell2003magnetism,chen2014anomalous}.

With the commensurate antiferromagnetic order, the analysis becomes significantly simplified due to the restored periodicity. The renormalized mean-field energy is given by,
\begin{equation}
\begin{aligned}
\tilde{\mathcal{E}}^{1d} =&  \sum_{j=1,2}\Big( \bar{\alpha}^{2}_{j} - \frac{1}{2}\sqrt{1+4g^{2} \bar{\alpha}^{2}_{j}} \Big)
+4\bar{J}\bar{\alpha}_{1}\bar{\alpha}_{2},
\end{aligned}
\label{Eq_1d_MF_2}
\end{equation}
where $j$ denotes the sublattice index, $\tilde{\mathcal{E}}^{1d}=2\langle H^{1d}\rangle /N_{a}N\Omega$, and $\bar{\alpha}_{j}=\sqrt{\omega/N_{a}\Omega }\langle a_{j,n}\rangle$ denotes the renormalized photon condensation at the odd $(j=1)$ and even $(j=2)$ sites. In this mean-field description, only two independent variables appear. The corresponding effective Hamiltonian describing quantum fluctuations is written as $\tilde{H}^{1d}_{g}=\sum^{\infty}_{n=-\infty} \tilde{\mathcal{H}}^{1d}_{n}$, where
\begin{equation}
\begin{aligned}
\tilde{\mathcal{H}}^{1d}_{n} =& \sum_{j=1,2} \Big[ \omega a^{\dagger}_{j,n}a_{j,n}+\Omega \sqrt{1+4g^{2}\bar{\alpha}^{2}_{j}} b^{\dagger}_{j,n}b_{j,n}  \\
&+ \lambda \frac{ \text{sgn}[\bar{\alpha}_{j}]}{\sqrt{1+4g^{2}\bar{\alpha}^{2}_{j}}} (a_{j,n}+a^{\dagger}_{j,n})(b_{j,n}+b^{\dagger}_{j,n})\Big]\\
&+ J(a^{\dagger}_{2,n}a_{1,n}+a^{\dagger}_{1,n+1}a_{2,n}+\text{H.c.} ) ,
\end{aligned}
\label{Eq_1d_Ham_3}
\end{equation}
where $\bar{\alpha}_{j}$, independent of the unit cell index $n$, is the solution that minimizes the renormalized mean-field energy described above. Because the system possesses translational symmetry with two sublattices per unit cell, the Hamiltonian density can be expressed as an $n$-independent (local) $8\times 8$ Bogoliubov-de Gennes Hamiltonian in momentum space. This structure enables the exact calculation of excitation spectra in the superradiant phase of the infinite simple Dicke chain model, which shows the standard excitation spectrum with a mean-field exponent $1/2$. On the other hand, the region near the domain wall must be treated separately from the commensurate bulk. Such domain-wall structures have been studied in spin systems, including antiferromagnets~\cite{jaramillo2007microscopic,kim2014propulsion,hedrich2021nanoscale}, and they are analogous to solitons~\cite{yoshimura2016soliton,meier2016observation}.

Our analysis in this section demonstrates that, for the Dicke chain, geometric frustration continues to influence the low-energy excitation spectrum at large odd-$N$ through an emergent Nambu-Goldstone mode near the critical point. Away from criticality, however, this influence fades as the domain wall localizes and the bulk properties become insensitive to frustration. These observations motivate a question: can one identify other lattice geometries that exhibit frustration independent of system size? We address the question in the next section by examining alternative lattice geometries.

\begin{figure*}[ht!]
    \centering
    \includegraphics[width=0.82\linewidth]{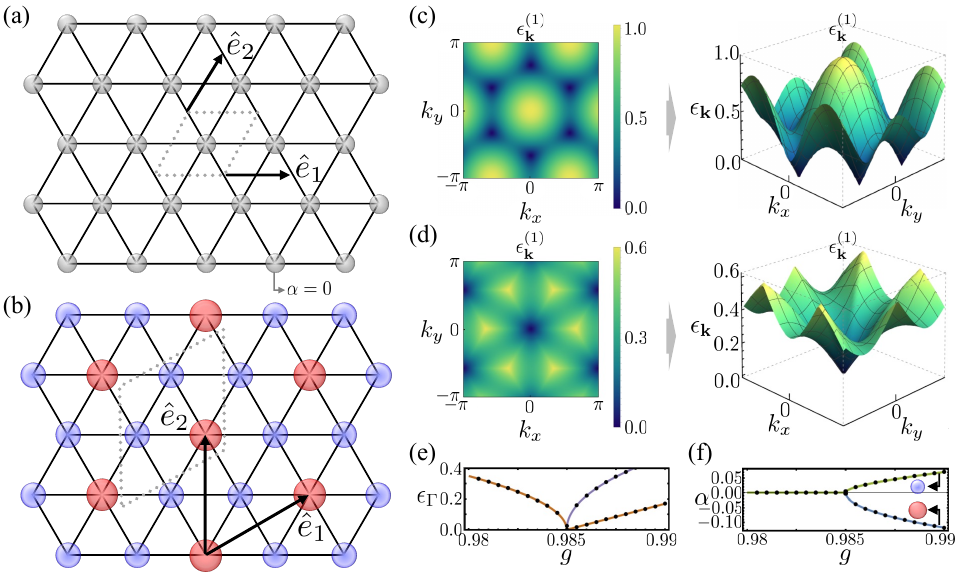}
    \caption{Schematics and numerical results for the infinite-sized Dicke triangular lattice model are shown. Schematic illustrations show the Dicke triangular lattice model in (a) the normal phase and (b) the superradiant phase, with circle size and color indicating the amplitude and sign of the order parameter, respectively. In (a), the gray dashed line and black arrows represent the minimal unit cell and the lattice vector of the triangular lattice. In (b), the gray dashed line and black arrows denote the extended unit cell, which includes three sublattices, along with the lattice vector. The excitation energy band is shown relative to the critical coupling strength, using (c) the minimal unit cell and (d) the extended unit cell in momentum space, where $\epsilon^{(1)}_{\bf k}$ denotes the lowest and next-lowest bands, respectively. (e) The excitation energy of the critical mode as a function of the effective coupling strength $g$. (f) The order parameter $\alpha$ for all sublattices as a function of the effective coupling strength $g$. In (e) and (f), the black dots indicate numerical data, while the colored lines are guides to the eye. The parameters are $\omega=10$, $\Omega/\omega=1$, and $\bar{J}=0.01$ for all data.}
    \label{fig_tri_1}
\end{figure*}

\section{Two-Dimensional Periodic Lattices}\label{Sec_2D}
We now extend the Dicke lattice model to a two-dimensional geometry. Building on the results of the one-dimensional model, one can deduce that Dicke models on bipartite lattices, such as square and honeycomb lattices, admit commensurate antiferromagnetic configurations for appropriate system sizes. Therefore, we consider the triangular lattice, where the non-bipartite structure is expected to introduce geometric frustration among the order parameters of neighboring sites; see Fig.~\ref{fig_tri_1}(a). This is motivated by the antiferromagnetic Ising model on the triangular lattice, where no state can simultaneously minimize all local terms of the Hamiltonian~\cite{wannier1950antiferromagnetism}, and the triangular unit cell, equivalent to a simple Dicke chain with $N=3$, which exhibits frustrated superradiance. We consider the nearest-neighbor hopping of the cavity photon where $J_{\mathbf{r},\mathbf{r}'}=J$, and $\phi_{\mathbf{r},\mathbf{r}'}=0$ for simplicity, so that the order parameters of the nearest neighbors prefer opposite directions. Its Hamiltonian is given by,
\begin{equation}
\begin{aligned}
H^{\text{tri}}=& \sum_{\mathbf{r}} \Big[ \omega a^{\dagger}_{\mathbf{r}}a_{\mathbf{r}} + \Omega J^{z}_{\mathbf{r}} 
+\frac{2\lambda}{\sqrt{N_{a}}}(a_{\mathbf{r}}+a^{\dagger}_{\mathbf{r}})J^{x}_{\mathbf{r}} \Big] \\
&+J \sum_{\mathbf{r},\mathbf{b}} (a^{\dagger}_{\mathbf{r}+\mathbf{b}}a_{\mathbf{r}} +a_{\mathbf{r}+\mathbf{b}}a^{\dagger}_{\mathbf{r}}) ,
\end{aligned}
\label{Eq_tri_Ham_1}
\end{equation}
where $\mathbf{r}$ denotes the position of the unit cell on the triangular lattice, 
\begin{equation}
   \mathbf{b}= \left\{ \hat{e}_{1},\: \hat{e}_{2},\: \hat{e}_{2}-\hat{e}_{1} \right\}
\end{equation}
with $\hat{e}_{1,2}$ being the lattice vector [Fig.~\ref{fig_tri_1}(a)]. The $\mathbb{Z}_2$ mirror symmetry remains intact. The Hamiltonian preserves the translational symmetry of the triangular lattice, remaining invariant under the coordinate transformation $\mathbf{r} \rightarrow \mathbf{r} + \hat{e}_{j}$, where $j = 1, 2$, with periodic boundary conditions.

\subsection{Photonic density-wave order}
The Hamiltonian in Eq.~(\ref{Eq_tri_Ham_1}) for the two-dimensional triangular lattice is expected to exhibit spontaneous symmetry breaking of both the parity and translational symmetry in the ground state upon increasing the local qubit-oscillator coupling, as observed in the one-dimensional model. Without prior knowledge of how translational symmetry is broken, whether partially or completely, determining the order parameter configuration would, in principle, require numerically minimizing an energy functional for an infinite system with infinitely many degrees of freedom, making a direct solution infeasible. In one dimension, we have established that the crystal momentum of the critical mode in the normal phase fixes the spatial periodicity of the superradiant order near the critical point. Here, we leverage this principle to determine the resulting translational symmetry of the superradiant phase in the triangular lattice, if any.
 
In the normal phase, the effective Hamiltonian for the excitation is given by,
\begin{equation}
\begin{aligned}
H^{\text{tri}}_{\rm n}=& \sum_{\mathbf{r}} \Big[ \omega a^{\dagger}_{\mathbf{r}}a_{\mathbf{r}} + \Omega b^{\dagger}_{\bf r}b_{\bf r} 
+\lambda (a_{\mathbf{r}}+a^{\dagger}_{\mathbf{r}})(b_{\bf r}+b^{\dagger}_{\bf r})  \Big] \\
&+J \sum_{\mathbf{r},\mathbf{b}} (a_{\mathbf{r}}a^{\dagger}_{\mathbf{r}+\mathbf{b}} +a^{\dagger}_{\mathbf{r}}a_{\mathbf{r}+\mathbf{b}}),
\end{aligned}
\label{Eq_tri_Ham_n_1}
\end{equation}
where $b_{\bf r}$ is the annihilation operator that describes the quantized fluctuation of the collective spin along the classical ground state. Since the normal phase Hamiltonian preserves the translational symmetry of the triangular lattice, it can be rewritten as a local Hamiltonian in momentum space through a Fourier transformation. We obtain the dispersion relation, namely the excitation energy in momentum space, as
\begin{equation}
    \epsilon_{\bf k} = \sqrt{ \frac{ \omega'^{2}_{\bf k}+\Omega^{2} \pm \sqrt{ (\omega'^{2}_{\bf k}-\Omega^{2})^{2}+16\omega'_{\bf k}\Omega \lambda^{2} } }{2} },
\end{equation}
where 
\begin{equation}
\omega'_{\bf k} = \omega+ 2J\sum_{\bf b} \cos(\bf k\cdot b).
\end{equation}
The onset of the superradiant phase occurs when the excitation gap closes, and we aim to identify the critical momentum, namely the momenta at which this closing occurs, as shown in Fig.~\ref{fig_tri_1}. Note that the stability of the normal phase requires that the following inequality holds for all momenta,
\begin{equation}
    \omega'_{\bf k}\ge \frac{4\lambda^{2}}{\Omega}.
\end{equation}
As the coupling strength $\lambda$ is increased, the minimum of the left-hand side approaches the right-hand side, signaling the onset of the gap closing. Consequently, the critical momentum is determined by the momenta at which the left-hand side attains its minimum. These critical momenta are given by,
\begin{equation}
    \mathbf{k}_{c} = \Big(\pm \frac{4\pi}{3},0\Big),\:\Big(\pm \frac{2\pi}{3},\frac{2\pi}{\sqrt{3}}\Big),\:\Big(\pm \frac{2\pi}{3},- \frac{2\pi}{\sqrt{3}}\Big),
\end{equation}
and one can verify that the excitation gap closes at these points in Fig.~\ref{fig_tri_1}(c). Thus, the modes corresponding to these momenta become critical and condense macroscopically in the superradiant phase.

The reciprocal lattice vectors in the triangular lattice are given by,
\begin{equation}
    \mathbf{G}_{1}= \Big(2\pi,-\frac{2\pi}{\sqrt{3}}\Big),\: 
    \mathbf{G}_{2}= \Big(0,\frac{4\pi}{\sqrt{3}}\Big).
\end{equation} The critical momentum vectors and reciprocal lattice vectors suggest a reduction in translational symmetry in the superradiant phase. By comparing the lengths of the reciprocal lattice vector $(\mathbf{G}_{j})$ and the critical momentum, i.e., 
\begin{equation}
|\mathbf{G}_{j}|/|\mathbf{k}_{c}|=\sqrt{3},
\end{equation}
we find that the length of the lattice vector and the area of the minimal unit cell extend by factors of $\sqrt{3}$ and $3$ in the superradiant phase, respectively. Therefore, we infer a reduced translational symmetry with three sublattices per unit cell in the superradiant phase, as illustrated in Fig.~\ref{fig_tri_1}(b). In the following, we verify the above prediction for the reduced translational symmetry by examining the excitation energy and the order parameter as a function of the coupling strength across the superradiant phase transition; see Appendix~\ref{Appx_Failure}.

Having identified the reduced translational symmetry in the superradiant phase, one can dramatically reduce the number of parameters appearing in the mean-field energy from $N$ to $3$: $\alpha_{\mathbf{r}},\: \theta_{\bf r},\: \phi_{\bf r}\rightarrow \alpha_{j},\: \theta_{j},\: \phi_{j}$ where $j=1,2,3$ is the sublattice index. Then, the renormalized mean-field energy with the reduced degrees of freedom is given by,
\begin{equation}
\begin{aligned}
\tilde{\mathcal{E}}^{\rm tri} =& \sum_{j=1,2,3}\Big[ \bar{\alpha}^{2}_{j}+ \frac{1}{2}\cos\theta_{j} + g \bar{\alpha}_{j} \sin\theta_{j}\cos\phi_{j}\Big] \\
&+6\bar{J} \Big[ \bar{\alpha}_{1}\bar{\alpha}_{2}+\bar{\alpha}_{2}\bar{\alpha}_{3}+\bar{\alpha}_{3}\bar{\alpha}_{1}  \Big],
\end{aligned}
\label{Eq_Tri_MF_2}
\end{equation}
where $\tilde{\mathcal{E}}^{\rm tri} = 3\langle H^{\rm tri} \rangle /N_{a}N\Omega $ and $N$ is the number of sites in the triangular lattice. The above mean-field energy can be readily minimized to find the order parameters, $\bar{\alpha}_{j}$ [See Fig.~\ref{fig_tri_1}].

The main finding of the analysis above is that the order parameter configuration of the extended triangular unit cell becomes identical to that of the Dicke triangle~\cite{zhao2022frustrated,zhao2023anomalous,luo2026quantum,zhang2024closed}, as illustrated in Fig.~\ref{fig_tri_1}(d). In other words, geometric frustration remains present within each extended unit cell, because the three antiferromagnetic photon-hopping bonds cannot be simultaneously minimized. However, once the system selects one of the six degenerate frustrated configurations at the unit-cell level, the same pattern can repeat throughout the lattice without introducing any additional frustration, so that the global degeneracy remains limited to six. As a result, the global state preserves translational symmetry that is different from the underlying lattice symmetry across the 2D lattice.

This reduction of translational symmetry is analogous to that in commensurate charge-density-wave phases in electronic systems~\cite{wilson1975charge,modesti2007insulating,li2013magnetic}, where the emergence of a periodic charge modulation reduces the translational symmetry of the underlying lattice and defines a larger real-space unit cell. In the present case, a similar reduction occurs, but it is the periodic modulation of the photon condensations that carries the broken translational symmetry. Moreover, the mechanism underlying this \emph{photonic density-wave order} is distinct from that of a conventional charge-density wave: it arises from geometric frustration in both the relative signs of the photon condensates and the accompanying amplitude modulation, rather than from an electronic instability tied to Fermi-surface nesting, electron-phonon, or electron-electron interactions~\cite{overhauser1962spin,rice1975new,gruner1988the,johannes2008fermi}. We emphasize that the amplitude modulation of the local photon condensation plays a crucial role in preventing a complete breakdown of translational symmetry, in contrast to the situation in an antiferromagnetic Ising model, where only the sign of the local spins is the local degree of freedom with a fixed spin angular momentum~\cite{wannier1950antiferromagnetism}.

\subsection{Robust anomalous critical scaling}
We now examine the nature of elementary excitations in the photonic density-wave ordered phase. By leveraging the mean-field solutions with reduced translational symmetry, we aim to obtain the band structure in the folded Brillouin zone associated with the enlarged real-space unit cell, including three bosons for cavities and three bosons for spins. To this end, we consider the quantum fluctuation around the mean-field solutions and get an effective Hamiltonian for the superradiant phase
\begin{equation}
    H^{\text{tri}}_{\rm g} = \sum_{\mathbf{r},j} \mathcal{H}^{\text{tri}}_{\mathbf{r},j} + J \sum_{\mathbf{r}} \mathcal{H}^{\text{tri}}_{1,\mathbf{r}},
\label{Eq_tri_Ham_s_1}
\end{equation}
with
\begin{equation}
\begin{aligned}
\mathcal{H}^{\text{tri}}_{\mathbf{r},j} =& \omega a^{\dagger}_{\mathbf{r},j}a_{\mathbf{r},j} + \Omega \sqrt{1+4g^{2}\bar{\alpha}^{2}_{j}} b^{\dagger}_{\mathbf{r},j}b_{\mathbf{r},j} \\
&+\lambda \frac{\text{sgn}[\bar{\alpha}_{j}]}{\sqrt{1+4g^{2}\bar{\alpha}^{2}_{j}}} (a_{\mathbf{r},j}+a^{\dagger}_{\mathbf{r},j})(b_{\mathbf{r},j}+b^{\dagger}_{\mathbf{r},j})  ,
\end{aligned}
\end{equation}
and
\begin{equation}
\begin{aligned}
\mathcal{H}^{\text{tri}}_{1,\mathbf{r}} =&  a^{\dagger}_{\mathbf{r},1}a_{\mathbf{r},2}+a^{\dagger}_{\mathbf{r},2}a_{\mathbf{r},3}+a^{\dagger}_{\mathbf{r},3}a_{\mathbf{r},1}+a^{\dagger}_{\mathbf{r},1}a_{\mathbf{r}+\hat{e}_{1},2}\\
&+a^{\dagger}_{\mathbf{r},1}a_{\mathbf{r}+\hat{e}_{2},3}
+a^{\dagger}_{\mathbf{r},1}a_{\mathbf{r}+\hat{e}_{2},2}+a^{\dagger}_{\mathbf{r},3}a_{\mathbf{r}+\hat{e}_{1},2}\\
&+a^{\dagger}_{\mathbf{r},1}a_{\mathbf{r}+\hat{e}_{2}-\hat{e}_{1},3}
+a^{\dagger}_{\mathbf{r},2}a_{\mathbf{r}+\hat{e}_{2}-\hat{e}_{1},3} +\text{H.c.},
\end{aligned}
\end{equation}
and $\bar{\alpha}_{j}$ is the order parameter that minimizes the mean-field energy of Eq.~(\ref{Eq_Tri_MF_2}). Here, $\mathbf{r}$ represents the position of the extended unit cell in Fig.~\ref{fig_tri_1}(b), so the Hamiltonian exhibits a reduced translational symmetry. This residual translational symmetry allows the Hamiltonian to be rewritten as a local Hamiltonian in momentum space through a Fourier transformation.

The resulting energy band structure in the photonic density-wave ordered phase is given in Fig.~\ref{fig_tri_1}(d), where only the lowest-energy band near the critical point is plotted. We find that the energy gap closes at the $\Gamma$-point, i.e., the critical momentum is $\mathbf{k}_{c} = (0, 0)$. This demonstrates that there is no further enlargement of the real-space unit cell, which is consistent with our prediction that the superradiant phase does indeed have the three-sublattice periodicity as we predicted in the previous section. We also plot the excitation energy of the critical mode at the $\Gamma$-point $(\epsilon_{\Gamma})$ as a function of the coupling strength in Fig.~\ref{fig_tri_1}(e). On both sides of the critical point, one of the bands exhibits a gap closing that follows a square-root law,
\begin{equation}
    \epsilon_{\Gamma}^\textrm{MF} \propto |g-g_c|^{\frac{1}{2}}.
\end{equation}
where the superscript MF indicates that this critical property is a characteristic of mean-field type phase transitions. Importantly, the additional band exhibiting a linear gap closing, namely,
\begin{equation}
    \epsilon_{\Gamma}^\textrm{F} \propto |g-g_c|^{1},
\end{equation}
persists in the infinite lattice limit. We note that this anomalous critical scaling of additional band $\epsilon_{\Gamma}^\textrm{F}$ is identical to that found in the Dicke triangle, indicating that the universal properties of the photonic density-wave order are entirely determined by geometric frustration within the enlarged triangular unit cell. This behavior stands in stark contrast to the one-dimensional chain studied in the previous section, where the linearly closing gap ultimately evolves into a Goldstone mode as the lattice size increases.

As a final remark, we note that the above analysis assumes boundary conditions chosen such that the entire system remains commensurate with the reduced periodicity imposed by the order-parameter configuration. For generic finite lattices, however, the imposed boundary conditions may become incompatible with the periodicity of the order parameter configuration, leading to a complete breakdown of translational symmetry. In such cases, we expect an emergent Nambu-Goldstone mode to appear below the square-root excitation branch near the critical coupling, analogous to the behavior found in the simple Dicke chain due to domain-wall formation, whose size scales with $O(N)$. Furthermore, for these generic system sizes and sufficiently far from the critical point, the domain-wall region is expected to become narrow, while the bulk of the system approaches a nearly commensurate ordered state, again mirroring the behavior of the simple Dicke chain. In this regime, we expect the Goldstone mode to become massive.

\section{Quasi-Periodic Phases} \label{Sec_Quasi} 
We have so far investigated the spontaneous breaking of translational symmetry and the nature of the resulting elementary excitations in the superradiant phase of Dicke lattice models, including the one-dimensional simple chain and the two-dimensional triangular lattice. In these models, a commensurate order parameter configuration can arise either in the thermodynamic limit of infinite lattice size or in a finite lattice of suitably chosen size. A commensurate configuration refers to a situation in which the emergent periodicity of the order parameter is commensurate with that of the underlying lattice, leading to a reduced translational symmetry in the system. This symmetry reduces the complexity of the governing equations and the number of independent variables, making the superradiant phase amenable to analytical treatment. The findings we have made so far therefore compel us to ask whether the superradiant phase can support qualitatively new forms of emergent periodicity that are incommensurate with the underlying lattice, and, crucially, how such ordering can be identified, and analyzed in the absence of a reduced translational symmetry or a commensurate unit cell. In this section, we address this question explicitly by analyzing the Dicke ladder, equivalently formulated as the $J_1-J_2$ Dicke chain, which provides a minimal setting for realizing and characterizing incommensurate quasi-periodic order.

\begin{figure*}[t!]
    \centering
    \includegraphics[width=0.90\linewidth]{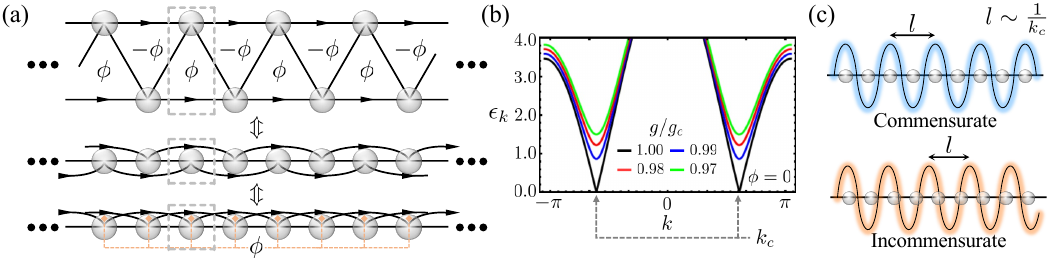}
    \caption{Schematic and numerical results for the Dicke ladder model. (a) Three equivalent representations of the Dicke ladder model, where each gray circle corresponds to a Dicke model. The black arrow indicates the gauge field direction of the magnetic flux. (b) Excitation energy $\epsilon_k$ plotted in momentum space for various effective coupling strengths $g$. Here, the parameters are $\omega=10$, $\Omega/\omega=1$, and $\bar{J}=0.01$. (c) Schematic illustrations showing cases where the order parameter configuration is commensurate (top) and incommensurate (bottom) with the lattice structure. 
    }
    \label{fig_ladder_1}
\end{figure*}

The Dicke ladder model is schematically illustrated in Fig.~\ref{fig_ladder_1}(a) (top)~\cite{li2023observation,halati2025exploring}. As a quasi-one-dimensional chain, the ladder represents an intermediate case between the simple Dicke chain model and the Dicke triangular lattice model. In this system, each triangular unit introduces frustration into the order parameter. In addition, we introduce an alternating magnetic flux threading each plaquette, breaking the time-reversal symmetry. It is illuminating to recognize that the Dicke ladder model with the nearest-neighborhood interaction is equivalent to the Dicke chain model with both the nearest-neighborhood and next-nearest-neighborhood interaction, which we refer to as the $J_1$-$J_2$ Dicke chain model [See Fig.~\ref{fig_ladder_1} (a) (middle)]. The alternating magnetic flux within each triangle arises from a uniform Peierls phase for the next-nearest-neighbor hopping, thus preserving the translational symmetry [See Fig.~\ref{fig_ladder_1}(a) (bottom)]. Consequently, the minimal unit cell of the Dicke ladder model contains a single sublattice, and the Hamiltonian reads
\begin{equation}
\begin{aligned}
H^{\text{lad}}=& \sum^{N}_{n=1} \Big[ \omega a^{\dagger}_{n}a_{n} + \Omega J^{z}_{n} 
+\frac{2\lambda}{\sqrt{N_{a}}}(a_{n}+a^{\dagger}_{n})J^{x}_{n} \\
&+J (a_{n}a^{\dagger}_{n+1} +e^{i\phi} a_{n}a^{\dagger}_{n+2} +\text{H.c.}) \Big] ,
\end{aligned}
\label{Eq_ladder_Ham_1}
\end{equation}
where $n$ indicates the position of the unit cell. Operators and parameters are defined consistently with those in the previous sections, including the periodic boundary condition. The Hamiltonian preserves the $\mathbb{Z}_{2}$ mirror symmetry and the translational symmetry of the quasi-one-dimensional chain, remaining invariant under the coordinate transformation $n \rightarrow n+1$ with periodic boundary conditions. The corresponding mean-field energy is given by
\begin{equation}
\begin{aligned}
\bar{\mathcal{E}}^{\rm lad} =& \sum^{N}_{n=1}\Big[ \bar{x}^{2}_{n}+\bar{y}^{2}_{n}+ \frac{1}{2}\cos\theta_{n} + g \bar{x}_{n} \sin\theta_{n}\cos\phi_{n} \\
&+2\bar{J} (\bar{x}_{n}\bar{x}_{n+1}+\bar{y}_{n}\bar{y}_{n+1}+\cos{\phi}(\bar{x}_{n}\bar{x}_{n+2}\\
&+\bar{y}_{n}\bar{y}_{n+2}) + \sin\phi (\bar{x}_{n}\bar{y}_{n+2}-\bar{x}_{n+2}\bar{y}_{n})  \Big],
\end{aligned}
\label{Eq_Ladder_MF_1}
\end{equation} 
where $\bar{\mathcal{E}}^{\rm lad} = \langle H^{\rm lad} \rangle / N_a \Omega$, $\bar{\alpha}_{n}=\bar{x}_{n}+i \bar{y}_{n}$, and $x_{n},\: y_{n}\in \mathbb{R}$. One can calculate the mean-field solutions of $\bar{\alpha}_n$, $\theta_n$, and $\phi_n$ that minimize the mean-field energy. Using these solutions, we expand the Hamiltonian and derive the effective Hamiltonian for the excitation as follows.
\begin{equation}
\begin{aligned}
H^{\text{lad}}_{\rm g}=& \sum^{N}_{n=1} \Big[ \omega a^{\dagger}_{n}a_{n} + \Omega \sqrt{1+4g^{2}\bar{x}^{2}_{n}} b^{\dagger}_{n}b_{n} \\
&+\lambda \frac{\text{sgn}[\bar{x}_{n}]}{\sqrt{1+4g^{2}\bar{x}^{2}_{n}}} (a_{n}+a^{\dagger}_{n})(b_{n}+b^{\dagger}_{n})  \\
&+J (a^{\dagger}_{n}a_{n+1}+e^{i\phi}a^{\dagger}_{n}a_{n+2}+\text{H.c.})
\Big],
\end{aligned}
\label{Eq_Ladder_Ham_General_1}
\end{equation}
where $\bar{x}_{n}$ is the parameter that minimizes the mean-field energy of Eq.~(\ref{Eq_Ladder_MF_1}) and $b_{n}$ is the annihilation operator that describes the quantized fluctuation of the collective spin. We note that this Hamiltonian covers both the normal phase ($\bar{\alpha}_{n}=0$) and the superradiant phase ($\bar{\alpha}_{n}\neq 0$).

\subsection{Irrational critical momentum}
To predict the emergent periodicity in the superradiant phase, we investigate the excitation energy spectrum in momentum space within the normal phase. In the absence of the spontaneous symmetry breaking, the model possesses the periodicity with a single sublattice per unit cell, regardless of the magnetic flux; see Fig.~\ref{fig_ladder_1}(a). Consequently, in momentum space, the Bogoliubov-de-Gennes Hamiltonian density is simply expressed as
\begin{equation}
    \mathcal{H}^{\text{lad,n}}_{k} = \frac{1}{2}\Psi^{\dagger}_{k} \mathcal{M}_{k} \Psi_{k},
\end{equation}
where the boson operator $\Psi_{k}$ and the matrix $\mathcal{M}_{k}$ in the Nambu spinor are given by,
\begin{equation}
\begin{aligned}
\Psi_{k}&=(a_{k},\: b_{k},\: a^{\dagger}_{-k},\: b^{\dagger}_{-k})^{\rm T},\\
    \mathcal{M}_{k}&= 
    \begin{pmatrix}
        \omega_{k} & \lambda & 0 & \lambda \\
        \lambda & \Omega & \lambda & 0 \\
        0 & \lambda & \omega_{-k} & \lambda \\
        \lambda & 0 & \lambda & \Omega 
    \end{pmatrix},
\end{aligned}
\end{equation}
where the bare dispersion relation of the cavity photons
\begin{equation}
 \omega_{k}= \omega+2J(\cos k +\cos{(2k+\phi)}).   
\end{equation}
The energy of the eigenmode is given by positive eigenvalues of the $4\times 4$ matrix $\mathcal{J}\mathcal{M}_{k}$ where $\mathcal{J}=\text{diag}(1,1,-1,-1)$.

We first consider the zero magnetic flux ($\phi = 0$) case. The excitation energy of the hybridized photons and qubits is given by,
\begin{equation}
    \epsilon^{\pm}_{k}=\sqrt{\frac{\omega^{2}_{k}+\Omega^{2}\pm \sqrt{(\omega^{2}_{k}-\Omega^{2})^{2}+16\omega_{k}\Omega\lambda^{2}}}{2}},
\end{equation}
and we plot the energy in momentum space for various coupling strengths in Fig.~\ref{fig_ladder_1}(b). Notably, the gap closes at a momentum $k_c$, which is not a rational multiple of $\pm \pi$ within numerical precision. Furthermore, the momentum of the critical mode remains constant despite changes in parameters such as $\Omega$ and $J$, although the critical coupling strength $\lambda_c$ varies. The non-zero value of the critical momentum, regardless of its exact value, indicates the spontaneous breaking of translational symmetry in the superradiant phase.  

We now determine the exact value of the critical momentum $k_{c}$ without the magnetic flux. In the normal phase, the following inequality holds for all momenta due to the positive semi-definite energy gap.
\begin{equation}
    \omega+2J(\cos k +\cos{2k}) \ge \frac{4\lambda^{2}}{\Omega}.
\end{equation}
Upon increasing the coupling strength $\lambda$, one may find a critical $\lambda_{c}$ where the equality holds at critical momentum $k_{c}$. Thus, minimizing the left-hand side yields the critical momentum,
\begin{equation}
\begin{aligned}
    k_{c} = \pm 2\tan^{-1}\sqrt{\frac{5}{3}}\neq 2\pi\frac{m}{n}\quad\text{for any } m,n \in \mathbb{Z}.
\end{aligned}
\label{Eq_kc_1}
\end{equation}
Remarkably, since $k_c/2\pi$ is irrational, the wavelength $2\pi/k_c$ associated with the critical mode is not commensurate with the underlying lattice, implying an incommensurate modulation in real space. Moreover, the value of $k_c$ is independent of the system size, indicating that it arises from the system's geometric structure and it is insensitive to the even-odd parity effects that arise in finite $1$D Dicke chains (see Eq.~\ref{Eq_discrete_k_1} for comparison). The effective critical coupling strength at which the critical mode closes the energy gap is 
\begin{equation}
\begin{aligned}
    g_{c} = \sqrt{ 1-\frac{9}{4}\bar{J} }. %\simeq \pm 1.8235,  
\end{aligned}
\label{Eq_lc_1}
\end{equation}
Both $k_c$ and $g_c$ predicted here are confirmed by the numerical results of Fig.~\ref{fig_ladder_1}(b).

Turning on the magnetic flux ($\phi \neq 0$), the energy band becomes asymmetric, and there is only one gap closing point at the critical condition due to time-reversal symmetry breaking, with the sign of the critical momentum depending on the direction of the flux. As the flux increases, the gap-closing point shifts towards $k = \pm \pi$ in momentum space. Namely, it is possible to tune the critical wave number between irrational and rational values by modulating the magnetic flux, which has a profound effect on the way the spontaneous symmetry breaking of the translational symmetry occurs.

\subsection{Incommensurate photonic density wave order}
From the photon condensation occurring at the irrational critical momentum $k_c$, we predict the emergence of quasi-periodic order since the periodicity of the order parameter configuration is incommensurate with the periodicity of the underlying lattice. We refer to this novel superradiant phase as the incommensurate photonic wave order, which is schematically illustrated in Fig.~\ref{fig_ladder_1}(c). Since no finite lattice translation leaves the order parameter invariant, the incommensurate photonic wave order is characterized by the complete spontaneous breaking of translational symmetry. Such quasiperiodic ordering has attracted considerable attention in condensed-matter physics, with prominent examples including quasicrystalline phases in twisted bilayer graphene~\cite{yao2018quasicrystalline,moon2019quasicrystalline} and incommensurate charge-density-wave phases~\cite{bao1993incommensurate,samnakay2015zone,miao2019formation}, while remaining largely unexplored in the context of superradiant phase transitions.

The quasiperiodicity implied by the irrational critical momentum makes a direct real-space analysis of the infinite ladder in the superradiant phase prohibitively difficult, as the absence of translational symmetry leads to a minimization problem involving infinitely many degrees of freedom. Therefore, we perform finite-size lattice calculations on periodic ladders to probe how the quasi-periodic ordering is realized as the system size increases, albeit with finite-size effects arising from the discretization of momentum. In doing so, the ansatz used to predict the emergent periodicity in the simple Dicke chain model in the previous section (Sec.~\ref{Sec_1D}) offers a useful insight. Since the Dicke ladder model is equivalent to a one-dimensional model with next-nearest-neighbor hopping, the order parameter at site $j$ in the superradiant phase in the vicinity of the critical point is expected to be
\begin{equation}
    \alpha_{j} = \alpha_{0} \cos(k^{*}j+\phi_{0}),
\label{Eq_alpha_ladder_1}
\end{equation}
where $1 \leq j \leq N$, and $k^*$ is a momentum that characterizes the periodicity. Here, $\alpha_0$ and $\phi_0$ are the amplitude and phase of the order parameter, respectively, assumed to be uniform across the sites. Under this ansatz, the minimization problem reduces to only two variables, $\alpha_0$ and $\phi_0$, regardless of the system size, which is significantly more efficient than bruteforce minimization of $N$ degrees of freedom. The detailed algorithm for finding the solution near and far from the critical point is presented in Appendix~\ref{Appx_Algorithm}.

\begin{figure}[t!]
    \centering
    \includegraphics[width=\linewidth]{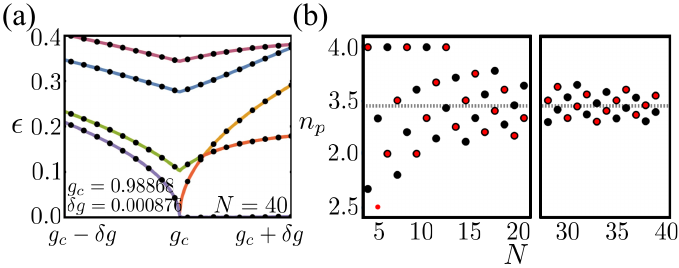}
    \caption{(a) Numerically computed excitation energy of the Dicke ladder model for system size $N = 40$ as a function of the coupling strength. Here, the black dots indicate numerical data, while the colored lines are guides to the eye. (b) Periodicity of the order parameter configuration $n_p$ in the superradiant phase as a function of the lattice size $N$. The black dots indicate the neighboring periodicity based on the allowed momentum neighboring the critical momentum $k_c$, while the red dots represent the numerical result. The gray dashed line indicates the periodicity corresponding to $k_{c}$. The parameters are $\omega=10$, $\Omega/\omega=1$, and $\bar{J}=0.01$ for all data.}
    \label{fig_ladder_3}
\end{figure}

We first perform this reduced minimization over the two parameters, and then refine the result by locally optimizing the full $N$-variable problem using the two-variable solution as a seed. This approach enabled us to investigate significantly larger systems. As a representative case, the excitation spectrum at $N = 40$ is shown in Fig.~\ref{fig_ladder_3}(a), which exhibits gap closing with the characteristic square-root behavior, on both sides of the transition. Importantly, there emerges an additional massless excitation for $g>g_c$ near the critical point. This Nambu-Goldstone mode originates from the spontaneous breaking of translational symmetry through the same mechanism as in the one-dimensional chain discussed in Sec.~\ref{Sec_1D}; see Appendix.~\ref{Appx_GoldstoneLadder} for a detailed discussion. It becomes massive as one moves far away from the critical point.

It is important to recognize that the value of $k^{*}$ is chosen from the set of allowed crystal momenta, which depend on the system size $N$. Upon changing the lattice size $N$, we numerically investigate the emergent periodicity of the superradiant phase. The emergent periodicity $n_{p} = 2\pi / k^{*}$ as a function of $N$ is presented in Fig.~\ref{fig_ladder_3}(b), where black dots represent periodicities corresponding to the allowed momenta neighboring to the critical momentum $k_c$ from Eq.~(\ref{Eq_kc_1}), while red dots indicate numerical results. The result demonstrates that the emergent periodicity $n_{p}$ tends to correspond to one of the allowed momenta neighboring $k_c$, and it progressively becomes closer to the irrational periodicity $2\pi/k_c$. This supports our prediction that, in the thermodynamic limit, the quasi-periodicity of the incommensurate photonic density wave order emerges.

\begin{table}[t!]
\begin{tabular}{ccc}
\hline \hline 
Periodicity \:\: & $k_{c}$\:\: & $\phi$  \\ \hline
2\:\:           &  $\pm \pi$\:\: & $0.5\pi \le \phi \le 1.5 \pi$ \\ \hline
3\:\:           &  $-2\pi/3$\:\: & $\simeq 0.32782\pi $ \\ \hline
4\:\:           &  $\pm \pi/2$\:\: & - \\ \hline
5\:\:          &  $-4\pi/5$ \:\: & $\simeq 0.40382\pi $ \\ \hline
\end{tabular}
\caption{Example of magnetic flux values for the commensurate order parameter configuration in the superradiant phase of the Dicke ladder model. The periodicity refers to the number of sublattices in a unit cell. }
\label{Table_Flux_1}
\end{table}

\subsection{Incommensurate-to-commensurate transition driven by magnetic flux}
The variation of the critical momentum with the magnetic flux suggests an intriguing possibility of inducing transitions between incommensurate and commensurate photonic density wave order by tuning the magnetic flux. We establish such a mechanism in this section. To this end, for varying coupling strength $\lambda$ and the magnetic flux $\phi$, we numerically investigate the critical momentum $k_c$ from the excitation energy band in the normal phase. We find that the zero-flux critical momentum in Eq.~(\ref{Eq_kc_1}) sets the lower bound for the allowed critical momentum for non-zero flux, namely, 
\begin{equation}
    2 \tan^{-1}\sqrt{\frac{5}{3}} \le |k_{c}| \le \pi.
\end{equation}
This inequality immediately rules out the possibility of a homogeneous order with a single site per unit cell for arbitrary values of the flux, since the critical momentum $k_c$ cannot vanish. Likewise, a commensurate order with four sites in an extended unit cell is excluded, as $k_c$ cannot take the values $\pm \pi/2$. In contrast, commensurate orders with emergent periodicities such as $2$, $3$, and $5$ remain allowed. For these cases, the corresponding magnetic flux values are determined numerically and are summarized in Table~\ref{Table_Flux_1}.

\begin{figure}[t!]
    \centering
    \includegraphics[width=\linewidth]{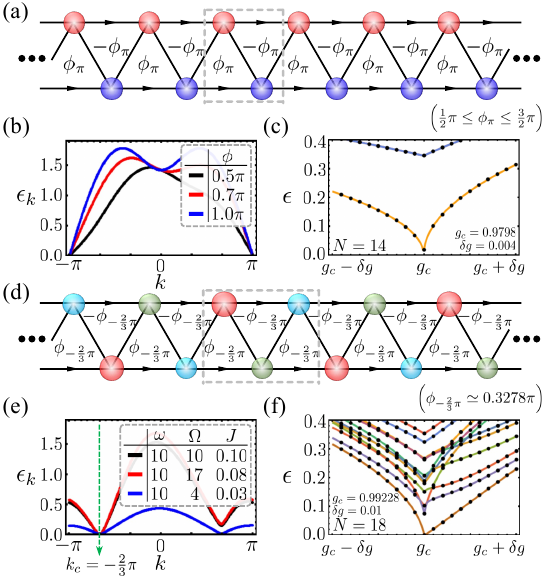}
    \caption{(a) Schematic of the Dicke ladder model, showing a unit cell with two sublattices and a magnetic flux $ \pm \phi_{\pi} $ associated with each triangle. (b) Excitation energy bands for different magnetic flux values $ \phi = 0.5\pi, \: 0.7\pi, \: \pi $ of the infinite ladder in momentum space. (c) The excitation energy of the ladder with $ N = 14 $ along the effective coupling strength $g$ for the magnetic flux $ \phi = \pi $. Here, the alternating order of (a) is enforced during the minimization of the mean-field energy. (d) Schematic of the Dicke ladder model, showing a unit cell with three sublattices and a magnetic flux $ \pm \phi_{-2\pi/3} $ associated with each triangle. (e) Excitation energy bands for different parameter conditions of the infinite ladder and the flux $\phi_{-2\pi/3}$ in momentum space. (f) The excitation energy of the ladder with $ N = 18 $ along the effective coupling strength $g$ for the magnetic flux $ \phi_{-2\pi/3} $. Here, the order of (d) is enforced during the minimization of the mean-field energy. In (c) and (f), the black dots indicate numerical data, while the colored lines are guides to the eye. The parameters are $\omega=10$, $\Omega/\omega=1$, and $\bar{J}=0.01$ for all data.}
    \label{fig_ladder_2}
\end{figure}

Next, we examine whether the magnetic flux values listed in Table~\ref{Table_Flux_1} yield the corresponding commensurate orders in the superradiant phase of the Dicke ladder model. We begin by considering the case of two sublattices per unit cell, as shown schematically in Fig.~\ref{fig_ladder_2}(a). The flux range from $0.5\pi$ to $1.5\pi$ corresponds to the critical momentum $\pm \pi$, as indicated by the excitation energy plot in the normal phase in Fig.~\ref{fig_ladder_2}(b). With $\phi = \pi$, we impose the alternating order parameter configuration and minimize the mean-field energy with two variables, respecting the periodicity. As a result, we observe that the excitation energy closes at the expected critical coupling strength and remains continuously connected across different coupling strengths. For illustration, we plot the case of periodic 14 sites in Fig.~\ref{fig_ladder_2}(c), which exhibits a square-root behavior of the excitation energy. 

We also explore the case of three sublattices per unit cell, as shown in Fig.~\ref{fig_ladder_2}(d). When $\phi = \phi_{-\frac{2\pi}{3}}$, as listed in Table.~\ref{Table_Flux_1}, the excitation energy in the normal phase consistently demonstrates a gap closing at $k = -\frac{2\pi}{3}$, regardless of the specific parameter conditions, as shown in Fig.~\ref{fig_ladder_2}(e). With this flux, we constrain the three-sublattice order parameter configuration and minimize the mean-field energy with three variables, again respecting the periodicity. The excitation energy closes at the expected critical coupling strength and remains continuously connected as the coupling strength changes. For this case, we plot the result for periodic $18$ sites in Fig.~\ref{fig_ladder_2}(f), which shows linear behavior of the excitation energy. We note that these results hold for larger systems qualitatively, including those with $N = 21$ and $N = 24$.

We note that the scaling behavior of the excitation energy in the superradiant phase differs between the two cases with distinct magnetic flux and periodicity, as shown in Figs.~\ref{fig_ladder_2} (c) and (f). In particular, for Fig.~\ref{fig_ladder_2}(f), the critical exponents on the two sides of the phase transition are different from each other. This behavior is reminiscent of the Dicke triangle in the presence of magnetic flux, where the interplay between time-reversal symmetry breaking and geometric frustration gives rise to anomalous critical scaling with a critical exponent of $3/2$~\cite{zhao2023anomalous}, which is different from the symmetric phase critical exponent of $1/2$. Overall, we emphasize that the magnetic flux values listed in Table.~\ref{Table_Flux_1} give rise to the emergence of commensurate photonic density wave order with the predicted periodicity in the superradiant phase of the Dicke ladder model.

\section{Extension to the quantum Rabi Model}\label{Sec_Rabi}
The quantum Rabi model (QRM), the single-atom limit of the Dicke model, is a paradigmatic model describing the coherent interaction between a bosonic mode (cavity photon) and a two-level system (such as a spin, atom, or qubit), governed
by the following Hamiltonian,
\begin{equation}
    H_{R} = \omega a^{\dagger}a + \frac{\Omega}{2}\sigma_{z} + \lambda (a+a^{\dagger})\sigma_{x},
\label{Eq_Ham_Rabi_1}
\end{equation}
where $a$ is the boson annihilation operator of the cavity photon, $\sigma_{x,y,z}$ is the Pauli matrix, $\omega$ and $\Omega$ denote frequencies of the two modes, and $\lambda$ denotes the coupling strength between them.

\begin{table*}[t!]
\begin{tabular}{lllll}
\hline \hline 
              & Conditions   & Anomalous scaling \:\:       & Degeneracy \:\: & Frustration \\ \hline
Simple Chain &  \begin{tabular}[c]{@{}l@{}} Commensurate order (even \# of sites) \\ Incommensurate order (odd \# of sites) \end{tabular} &  \begin{tabular}[c]{@{}l@{}}No\\ Yes \end{tabular}   & \begin{tabular}[c]{@{}l@{}}2\\ $O(N)$ \end{tabular}  & \begin{tabular}[c]{@{}l@{}}No\\ Yes \end{tabular}  \\ \hline
Triangular lattice\:\: & \begin{tabular}[c]{@{}l@{}} Commensurate order \\ Incommensurate order \end{tabular} & \begin{tabular}[c]{@{}l@{}}Yes (linear)\\ Yes \end{tabular} & \begin{tabular}[c]{@{}l@{}}3\\ $O(N)$ \end{tabular} & \begin{tabular}[c]{@{}l@{}}Yes\\ Yes \end{tabular}  \\ \hline
Ladder & \begin{tabular}[c]{@{}l@{}}General conditions\\ Special flux \end{tabular} & \begin{tabular}[c]{@{}l@{}}Yes \\ Yes (except $\phi_{\pi}$)\end{tabular} & \begin{tabular}[c]{@{}l@{}} $O(N)$  \\  \# of sublattices \end{tabular} & \begin{tabular}[c]{@{}l@{}}Yes \\ Yes \end{tabular} \\ \hline
\end{tabular}
\caption{Summary of ground state and excitation properties in the superradiant phase for various Dicke lattice models. The anomalous scaling refers to the critical behavior of the excitation energy as a function of the coupling strength that deviates from the typical square-root dependence observed in the superradiant phase. Degeneracy counts the number of energetically equivalent classical ground states. $N$ denotes the number of lattice sites in the system. }
\label{Table_summary_1}
\end{table*}

The QRM exhibits critical behavior belonging to the same universality class as the Dicke model in the limit $\Omega/\omega \rightarrow \infty$, which is analogous to the thermodynamic limit $N_a \rightarrow \infty$ in the Dicke model~\cite{hwang2015quantum}. In this regime, the QRM undergoes a second-order superradiant phase transition, characterized by macroscopic condensation of the bosonic mode in the ground state and spontaneous breaking of parity symmetry. Crucially for our discussion, the mean-field energy of the QRM in the $\Omega/\omega\rightarrow\infty$ limit becomes identical to that of the Dicke model in the $N_{a}\rightarrow \infty$ limit, once the atomic degrees of freedom of the Dicke model are integrated out and the photonic condensation $\alpha$ is renormalized as $\bar\alpha=\alpha/\sqrt{N_{a} \Omega/\omega}$~\cite{zhao2022frustrated}. Therefore, the mean-field solutions for the cavity photons are identical between the Dicke lattice models and the quantum Rabi lattice models, where the on-site Hamiltonian in Eq.~\eqref{Eq_H_loc_1} is replaced by Eq.~\eqref{Eq_Ham_Rabi_1}. Recent studies have explicitly demonstrated this correspondence between the QRMs and Dicke models in small-size lattice systems~\cite{zhao2022frustrated,zhao2023anomalous,zhang2021quantum,fallas2022understanding,li2023quantum}. In particular, the phase diagram and critical exponents in one-dimensional chains, both with and without flux insertion, show strong quantitative agreement for the Dicke lattice model~\cite{zhao2022frustrated,zhao2023anomalous} and the quantum Rabi lattice models~\cite{zhang2021quantum,fallas2022understanding,li2023quantum}.

This implies that all superradiant orders predicted for the Dicke lattice model in previous sections are expected to emerge in quantum Rabi lattice models with the same geometry. Consequently, our predictions extend directly beyond Dicke lattice models to quantum Rabi lattice models, which can be realized across a wide range of physical platforms — including superconducting circuits~\cite{Zheng2023SC}, trapped ions~\cite{cai2021observation}, and optomechanical systems~\cite{Wang2024Opto} — where accessing the superradiant phase is experimentally feasible. This correspondence substantially broadens the scope and experimental relevance of our results, and positions our work as a unifying framework for understanding frustrated superradiant order across distinct microscopic realizations that share the same effective mean-field energy structure.

\section{Physical Realization}\label{Sec_Realization}
The experimental realization of the novel frustrated phases of the Dicke/Rabi lattice model requires two key ingredients: the engineering of strong local spin-boson coupling sufficient to drive the superradiant phase transition at each individual site, and the construction of a bosonic lattice with the desired geometry, in which neighboring sites are coherently coupled. Remarkable experimental progress has been achieved on both fronts across a variety of quantum platforms composed of coupled bosonic modes and effective spins, thanks to recent advances in engineering large-scale quantum systems. In the following, we review these two lines of experimental progress in turn, and discuss how their combination can provide a viable route toward realizing the frustrated superradiant phases predicted in this work.

First, the Dicke model and the quantum Rabi model on a single site with strong coherent spin-boson coupling have been proposed and realized across a variety of quantum platforms. In trapped-ion systems, the internal electronic states of the ions are coupled to their motional degrees of freedom via bichromatic Raman laser fields~\cite{Monroe2021RMP}. The detuning and intensity of these fields enable precise control over both the type and strength of the spin-boson coupling, allowing for the implementation of both models~\cite{,cohn2018bang,safavi2018verification,cai2021observation,Zhao2025Ion,guo2026ion2D,bullock2026ion2D}. In particular, the superradiant phase has been observed in experiments with a single ion~\cite{cai2021observation,Zhao2025Ion}, by leveraging an alternative thermodynamic limit based on a large frequency ratio in a finite-component system. The recent realization of the Dicke model in two-dimensional ion crystals with over $100$ ions, on the other hand, demonstrates significant progress toward scaling to larger system sizes, enabling controlled access to the thermodynamic limit~\cite{guo2026ion2D,bullock2026ion2D}. In cavity and circuit QED systems, strong light-matter coupling has been achieved using both optical and microwave cavities. In the optical cavity setting, collections of atoms~\cite{zhang2017Dicke,zhang2018Dicke} or Bose-Einstein condensates~\cite{baumann2010dicke,klinder2015dynamical,kroeze2018spinor} act as effective two-level systems collectively coupled to a single cavity mode, where the superradiant phase has been realized. In the microwave cavity setting, either superconducting qubits such as transmons~\cite{Zheng2023SC} or solid-state emitters, such as nitrogen-vacancy centers in diamond~\cite{Angerer2018superradiant} or spin ensembles in magnetic materials~\cite{song2025single}, are coupled to on-chip microwave resonators in the circuit QED architecture. The former platform has been used to simulate the quantum Rabi model and its superradiant phase, while the latter realizes the Dicke model using many two-level systems. Additionally, further theoretical proposals have explored the use of hybrid quantum systems~\cite{marquez2024quantum,lee2023cavity,lee2025diverging,Wang2024Opto}.

Second, the realization of a bosonic lattice can be achieved using scalable architectures for quantum technologies, with trapped-ion systems and superconducting circuits being the most suitable platforms. In ion traps, the local motional modes of ions in an ion crystal can be used to construct a bosonic lattice system~\cite{Toyoda2013Ion, Debnath2018Ion,Ohira2021Ion, Mei2022Ion}, where hopping interactions arise from the Coulomb interaction with a controllable range. There has been an remarkable progress in creating a large-scale $2D$ ion crystals~\cite{Bollinger2012ion,Bollinger2016ion,qiao2024tunable,luo2026quantum,guo2024site} with a possibility of creating a traingular lattice structure~\cite{Bollinger2012ion,qiao2024tunable}, which have been used to simulate frustrated spin models~\cite{kim2010quantum,islam2013emergence,Bollinger2012ion,mei2022experimental}. By utilizing the local motional modes, these systems can also be used to realize $2D$ bosonic lattices. In superconducting circuit platforms, bosonic lattices can be engineered using arrays of coupled microwave resonators within the circuit QED architecture~\cite{houck2012chip,koch2010time}. Each resonator mode naturally serves as a bosonic degree of freedom, while photon hopping between sites can be realized through capacitive or inductive couplings, with strengths that are highly tunable through circuit design. Recent experiments have demonstrated one and two-dimensional lattices of superconducting resonators~\cite{fitzpatrick2017observation,sundaresan2019cqed,kollar2019hyperbolic, kim2021cqed,zhang2023cqed,ma2019mott} with programmable connectivity and controllable band structures, enabling the realization of a wide range of tight-binding models for microwave photons. Combined with the ability to engineer synthetic magnetic flux~\cite{roushan2017chiral,koch2010time} and to incorporate collections of solid-state emitters~\cite{zou2014implementation,lee2023cavity,lee2025diverging} that drive the local superradiant transition, superconducting resonator arrays offer a promising platform for exploring frustrated bosonic phases.

These experimental advances in realizing strong local spin-boson coupling, enabling superradiant phase transitions, and engineering scalable bosonic lattices together indicate that all key ingredients to explore the experimental realization of exotic bosonic phases in Dicke and Rabi lattice systems are now within reach. Their combination therefore provides a realistic and promising route toward such realizations, either in trapped-ion platforms or superconducting circuits.

\section{Conclusion}\label{Sec_Conclusion}
In summary, we have demonstrated that large arrays of the Dicke model on periodic lattices realize a new class of quantum many-body systems in which each lattice site possesses an intrinsic thermodynamic limit in addition to thermodynamic limit of the lattice size. The strong light-matter coupling induces a superradiant phase transition similar to that of the conventional Dicke model. When geometric frustration is present on the lattice, competition between neighboring order parameters drives spontaneous breaking of translational symmetry and leads to anomalous superradiant phases. A summary of our findings is presented in Table~\ref{Table_summary_1}.

We addressed the difficulty of characterizing the ground-state order parameters arising from the frustration and the associated breaking of translational symmetry in the superradiant phase of large lattices, analogous to the Ising antiferromagnet on the triangular lattice. Instead of performing computationally demanding mean-field minimization over extensive variables, we showed that the problem can be simplified by identifying the critical momenta from the excitation spectrum at the critical point. These momenta directly determine the broken or reduced translational symmetry in the superradiant phase, allowing efficient treatment even for infinitely large periodic lattices such as the triangular lattice.

One of the most prominent features of the frustrated superradiant phase, reported in numerous previous works~\cite{zhao2022frustrated,zhao2023anomalous,padilla2022understanding,li2023quantum,xu2024quantum,qin2024quantum,luo2026quantum,zhang2024closed,cheng2022chiral} on small lattice systems, is the presence of an anomalous critical mode whose excitation energy vanishes more rapidly than the mean-field square-root scaling. We show that this anomalous mode evolves asymptotically into a Nambu-Goldstone mode associated with the spontaneous breaking of translational symmetry in the thermodynamic limit. We refer to it as an emergent Nambu-Goldstone mode in two senses: it appears only near the critical coupling within the superradiant phase and only in the large-system limit. This establishes a novel mechanism to featuring gapless excitations in the absence of continuous symmetry. Observing the emergence of a Nambu-Goldstone mode asymptotically as a function of system size is unusual, as spontaneous symmetry breaking is typically defined directly in the thermodynamic limit of the system size. Away from the critical point, the anomalous mode becomes lifted, and the disparity between even and odd cases disappears, except for the presence of a sharp domain wall in the order parameter configuration. This indicates that the even-odd parity distinction persists in the infinite size limit at low energies, only in the vicinity of the critical point.

We showed that frustration induces both commensurate and incommensurate orders in the superradiant phase of large periodic lattices. In the triangular lattice, the original translational symmetry is spontaneously broken, but a reduced periodicity emerges in the ground state. We refer to this as a photonic density-wave order, in analogy with charge-density-wave order in condensed matter systems. In contrast, for the ladder geometry, the critical momenta indicate that the photon condensation is always incommensurate with the lattice, implying the absence of translational symmetry in the superradiant phase, irrespective of system size. Finally, we showed that the periodicity of the condensation can be tuned by the magnetic flux through the plaquette, and identified conditions under which translational symmetry is restored in the superradiant phase.

\section{Outlook}\label{Sec_Outlook}
Our study provides a novel platform for exploring the spontaneous breaking of translational symmetry, quantum critical phenomena, and collective excitations in various periodic lattices. We anticipate that symmetry breaking in the superradiant phase may enrich (higher-order) topological phases in models such as the Haldane model~\cite{haldane1988model}, the Su-Schrieffer-Heeger model~\cite{su1979solitons,su1980soliton}, and their variants~\cite{benalcazar2017quantized,schindler2018higher,ezawa2018higher}. In such systems, strong light-matter coupling could give rise to photonic localized states at the boundaries. We expect our work to bridge and extend the frontiers of condensed matter physics, cavity quantum electrodynamics, and quantum optics.

The softening of the critical mode associated with translational symmetry breaking is analogous to the Kohn anomaly observed in phonon spectra in condensed matter systems~\cite{kohn1959image}. The Kohn anomaly refers to the softening of phonons at specific points in momentum space due to strong electron-phonon interactions. This behavior indicates a structural instability and a transition into a charge-density-wave phase, which corresponds to the spontaneous breaking of translational symmetry~\cite{renker1973observation,piscanec2007optical,li2021observation}. Similar to our study, the critical momentum determines the periodicity of the symmetry-broken phase, as exemplified by the Peierls transition in one-dimensional systems~\cite{pouget2016peierls}.

We focused solely on a closed system, neglecting interactions with the environment. However, it is also possible to incorporate dissipation by treating the system as an open quantum system, since all real experimental systems inevitably interact with their surroundings~\cite{diehl2008quantum,eisert2015quantum}. Naively, one might expect different scaling behaviors, as suggested by studies of the conventional Dicke and Rabi models under dissipation~\cite{baumann2011exploring,brennecke2013real,hwang2018dissipative,kirton2019introduction}. Moreover, the broken translational symmetry may change in the strong coupling regime~\cite{zou2014implementation}, and introducing different types of dissipators could give rise to alternative phases, potentially both static and dynamical~\cite{fruchart2021non,zelle2024universal,belyansky2025phase}. Extending the present framework to handle large system sizes with broken periodicity in the context of open quantum systems remains an important direction for future work.

While our analysis has focused on short-range interactions, such as nearest-neighbor hopping, it is also possible to extend this framework to include long-range interactions. For example, in superconductors, the Nambu-Goldstone mode acquires a finite energy gap due to long-range Coulomb interactions, a phenomenon known as the Anderson-Higgs mechanism~\cite{anderson1963plasmons}. Similarly, the phason mode in the charge-density-wave phase may become massive through Coulomb interaction~\cite{virosztek1993collective,kim2023observation}. Analogously, incorporating long-range interactions into our Dicke lattice model may enhance frustration, potentially giving rise to novel phases and excitation spectra. This would allow for a detailed investigation of how the emergent Nambu-Goldstone mode evolves with both interaction range and system size.

We believe that the rich phenomenology described here is accessible in current experimental platforms and expect that future work will further clarify the interplay between frustration, topology, dissipation, and long-range interactions in cavity-coupled quantum matter.

\acknowledgments
J.M.L. acknowledges support from Quantum Horizons Alberta. M.-J.H. acknowledges support from the Innovation Program for Quantum Science and Technology 2021ZD0301602. We thank Joseph Maciejko, Canon Sun, and Min Ju Park for the discussions. J.M.L. thanks Hyun-Woo Lee for his support and encouragement during this work.\\

\appendix

\section{FAILURE OF THE PERIODICITY ESTIMATION} \label{Appx_Failure}
Determining the correct periodicity in the superradiant phase of the Dicke lattice model is challenging without analyzing the critical momentum. To demonstrate the consequences of an incorrect assumption, we present an example where an incorrect choice of periodicity leads to a wrong result. Specifically, we revisit the Dicke ladder model discussed in Sec.~\ref{Sec_Quasi}, now considered without magnetic flux. This example is provided solely for illustrative purposes and does not correspond to any physically meaningful solution.

We suppose one assumes that the superradiant phase adopts a two-sublattice periodicity within each unit cell, motivated by the apparent preference for antiferromagnetic ordering between neighboring sites as suggested by the effective mean-field energy in Eq.~(\ref{Eq_Ladder_MF_1}). Under this assumption, the mean-field energy can be reformulated by reducing the number of variables as follows.
\begin{equation}
\begin{aligned}
\bar{\mathcal{E}}^{\rm lad}_{2} =& \sum_{j=1,2}\Big[ \bar{\alpha}^{2}_{j}+ \frac{1}{2}\cos\theta_{j} + g \bar{\alpha}_{j} \sin\theta_{j}\cos\phi_{j} \Big]\\
&+2\bar{J} (2\bar{\alpha}_{1}\bar{\alpha}_{2}+\bar{\alpha}^{2}_{1}+\bar{\alpha}^{2}_{2}),
\end{aligned}
\label{Eq_Ladder_MF_Neel_1}
\end{equation} 
where $\bar{\mathcal{E}}^{\rm lad}_{2} = 2\langle H^{\rm lad} \rangle / N N_a \Omega $, and $j=1,2$ denotes the sublattice. Then, the effective Hamiltonian for the excitations can be expressed using a reduced set of order parameters, assuming a large or even infinite system.

\begin{figure}[t!]
    \centering
    \includegraphics[width=0.985\linewidth]{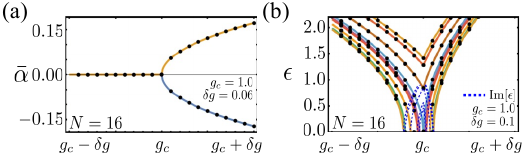}
    \caption{(a) Mean-field solution and (b) excitation spectrum of the Dicke ladder model with 16 sites, under an incorrect assumption of periodicity in the superradiant phase. The blue dashed line represents the imaginary part of the excitation energy, while the solid lines indicate the real parts. Note that these results are unphysical and demonstrate the failure arising from an incorrect periodicity assumption. The black dots indicate numerical data, while the colored lines are guides to the eye. The parameters are $\omega=10$, $\Omega/\omega=1$, and $\bar{J}=0.01$ for all data.}
    \label{fig_failure_1}
\end{figure}

In Fig.~\ref{fig_failure_1}, we present the mean-field solution and excitation energy as functions of the coupling strength, based on the above assumption and using a finite system of 16 sites. The plot shows apparent symmetry breaking and gap closing; however, these features are unphysical. Specifically, the excitation energy does not vanish at the effective critical coupling $g_c$, where the order parameter signals symmetry breaking. Instead, the energy gap closes at a different point and remains closed over a finite range, during which the excitation energy acquires an imaginary component (indicated by the blue dashed line). These signatures highlight the consequences of an incorrect periodicity assumption, and can be used to validate the correctness of the calculation.

\section{ALGORITHM FOR DETERMINING THE MEAN-FIELD SOLUTION IN THE SUPERRADIANT PHASE} \label{Appx_Algorithm}
In this appendix, we present the numerical procedure used to determine the mean-field solution and the corresponding excitation spectrum in the superradiant phase of a finite but large Dicke lattice. The key difficulty arises from the fact that, once translational symmetry is spontaneously broken, the periodicity of the order parameter is not known a priori. Consequently, the correct ordering pattern must be identified self-consistently near the critical point and then traced continuously as the coupling strength increases. 

\begin{algorithm}[H]
\caption{Mean-field solution in the superradiant phase}
\BlankLine

\textbf{Normal phase:} For $\lambda < \lambda_c$, the mean-field solution is the trivial configuration ($\alpha = 0$)\;

\textbf{Determine $\lambda_c$:} 
Compute the excitation spectrum in the normal phase and identify the coupling strength $\lambda_{c}$ where the lowest mode becomes critical\;

\textbf{Determine $k_c$:} 
Using the dispersion at $\lambda=\lambda_c$, identify the momentum $k_c$ where the gap closes\;

\textbf{Initialization near the critical point:} 
Choose $\lambda = \lambda_c + \epsilon$ (small $\epsilon > 0$)\;
Assume a periodicity determined by $k_c$, initialize the order parameter (amplitude and phase), and minimize the mean-field energy\;

\textbf{Check excitation continuity:} 
Compute the excitation spectrum\;
\If{the spectrum does not connect continuously to that at $\lambda_c$}{
  Consider periodicities from allowed discrete momenta neighboring $k_c$\;
  Repeat the initialization and minimization steps\;
}

\textbf{Continuation method:} 
Increase $\lambda$ slightly\;
Use the previous mean-field solution as the seed and minimize the mean-field energy again, ensuring continuity with respect to $\lambda$\;

\textbf{Iteration:} 
Repeat the continuation step to obtain the mean-field solution and excitation spectrum across the desired coupling range\;

\end{algorithm}

The algorithm above summarizes the steps used to (i) determine the critical coupling $\lambda_c$ from the normal-phase excitation spectrum, (ii) identify the critical momentum $k_c$ that predicts the emergent periodicity, and (iii) iteratively construct the mean-field solution throughout the superradiant phase by ensuring continuity of both the order parameter and the excitation spectrum.

\section{EMERGENT NAMBU-GOLDSTONE MODE IN THE DICKE LADDER MODEL}\label{Appx_GoldstoneLadder}
In Sec.~\ref{Sec_Quasi}, we numerically demonstrated that an emergent Nambu-Goldstone mode appears near the critical point in the large-$N$ limit. This mode is associated with the spontaneous breaking of translational symmetry, analogous to the one-dimensional chain discussed in Sec.~\ref{Sec_1D}. Here, we clarify the emergent continuous symmetry and the emergent Nambu-Goldstone mode in the superradiant phase of the Dicke ladder model.

We have calculated the critical momentum $k_{c}$ in the infinite-$N$ limit in Eq.~(\ref{Eq_kc_1}) for the Dicke ladder model in the normal phase. For a finite ladder of length $N$ with periodic boundary conditions, the allowed momenta are discrete, as shown in Eq.~(\ref{Eq_discrete_k_1}). The finite-size critical momentum is given by the allowed value of $k$ that lies close to $k_c$. Also, near the critical point in the superradiant phase, the order parameter can be written as,
\begin{equation}
    \alpha_j = \alpha_0 \cos \bigl(k^* j+\theta_0\bigr),
\label{Eq_alpha_appx_1}
\end{equation}
where $k^*$ sets the periodicity of the order parameter configuration. In general, $k^*$ coincides with a discrete momentum selected in the finite system and therefore does not exactly equal $k_{c}$, but approaches $k_c$ as $N$ increases, consistent with the main text and Fig.~\ref{fig_ladder_3}.

A lattice translation by $l$ sites ($\mathcal{T}_{l}$) acts as $j \rightarrow j+l$, which transforms Eq.~(\ref{Eq_alpha_appx_1}) as
\begin{equation}
    \mathcal{T}_{l}\alpha_{j}\equiv \alpha_{j+l}=\alpha_j(\theta+\delta \theta).
\end{equation}
where $\delta \theta =k^{*} l \quad (\mathrm{mod}\ 2\pi)$. Thus, translation is equivalent to shifting the phase and $l=1,\dots,N$ generates a discrete set of phase shifts for a finite system.

In the thermodynamic limit of $N$, the spacing between allowed phase shifts becomes dense, effectively restoring a continuous degree of freedom $\delta \theta \in[0,2\pi)$. For example, when $N=50$ the momentum spacing is of order $10^{-1}$, whereas for $N=10^{4}$ it is reduced to order $10^{-3}$. As a result, different order-parameter configurations are continuously connected by translation operations, which gives rise to the emergent Nambu-Goldstone mode near the critical point.

\bibliography{BibRef}

\end{document}